\pdfoutput=1
\documentclass[aj,numberedappendix]{emulateapj}

\newcommand\rmxaa{RMAA}
\newcommand{\kms}{\ensuremath{\mathrm{km\ s}^{-1}}}

\newcommand\htwo{H$_{\rm 2}$}
\newcommand\Halpha{H$\alpha$}

\newcommand\ori{$\theta ^{1}$Ori~C}
\newcommand\oriA{$\theta ^{2}$Ori~A}

\newcounter{ionstage}

\newcommand\Vt{\ensuremath{V_\mathrm{T}}}

\newcommand\Vsun{\ensuremath{V_\sun}}
\newcommand\Vomc{\ensuremath{V_\mathrm{OMC}}}

\slugcomment{Accepted for publication in the Astronomical Journal}

\shorttitle{\uppercase{High Velocity Features in the Orion Nebula}}
\shortauthors{\uppercase{O'Dell \& Henney}}

\begin{document}


 \title{High Velocity Features in the Orion Nebula\footnotemark[1,2]}

\author{C. R. O'Dell}
\affil{Department of Physics and Astronomy, Vanderbilt University, Box 1807-B, Nashville, TN 37235}

\and 

\author{W. J. Henney}
\affil{Centro de Radioastronom\'{\i}a y Astrof\'{\i}sica, Universidad Nacional Aut\'onoma de M\'exico, Apartado Postal 3-72,
58090 Morelia, Michaoac\'an, M\'exico}

\email{cr.odell@vanderbilt.edu}

\begin{abstract}
We have used widely spaced in time Hubble Space Telescope images to determine tangential velocities of features associated with outflows from young stars. These observations were supplemented by groundbased telescope spectroscopy and from the resultant radial velocities, space velocities were determined for many outflows. Numerous new moving features were found and grouped into known and newly assigned Herbig Haro objects. It was found that stellar outflow is highly discontinuous, as frequently is the case, with  long term gaps of a few hundred years and that these outflow periods are marked by staccato bursts over periods of about ten years. Although this has been observed in other regions, the Orion Nebula Cluster presents the richest display of this property. Most of the large scale Herbig Haro objects in the brightest part of the Orion Nebula appear to originate from a small region northeast of the strong Orion-S radio and infrared sources. With the possible exception of HH 203, we are not able to identify specific stellar sources, but do identify candidate sources for several other bright Herbig Haro objects. We find that there are optical features in the BN-KL region that can be related to the known large scale outflow that originates there. We find additional evidence for this outflow originating 500--1000 years ago.
\end{abstract}





\section{Introduction}
\footnotetext[1]{Based on observations with the NASA/ESA Hubble Space Telescope,
obtained at the Space Telescope Science Institute, which is operated by
the Association of Universities for Research in Astronomy, Inc., under
NASA Contract No. NAS 5-26555.}
\footnotetext[2]{Based on observations at the San Pedro Martir Observatory
    operated by the Universidad Nacional Aut\'onoma de M\'exico.}
\addtocounter{footnote}{2}
The long operational life of the Hubble Space Telescope (HST) has presented the opportunity to obtain and compare emission-line images of the central (Huygens) region of the Orion Nebula over a sufficiently long time-base to determine highly accurate motions in the plane of the sky.  These motions enable us to study the outflow that accompanies early phases of star formation and the interaction of this out-flowing plasma with the ambient nebular material.  An introduction to the Orion Nebula and its associated cluster is presented in several recent review articles \citep{ode01,mue08, ode08}.

In this paper we present the results of  the determination of new tangential and radial velocities and their interpretation.  We will show that a number of new stellar outflows (Herbig Haro objects, HH) have been found, that some previously designated HH objects are not what was originally argued, and that these motions provide us with important new information about star formation in the visually obscured BN-KL and Orion-S centers. 

\section{Observations}\label{obs}

Our observational material consists of both new spectroscopic and imaging data. The spectroscopic
observations were made at Observatorio Astron\'omico Nacional en San Pedro M\'artir operated by the Universidad Nacional Aut\'onoma de M\'exico using the MEZCAL echelle spectrograph \citep{mea03}. The new imaging observations were made with the HST's WFPC2 and ACS cameras as part of program GO 10921.  The data were processed with the IRAF package of software.\footnote{IRAF is distributed by the National Optical Astronomy Observatories, which is operated by the Association of Universities for Research in Astronomy, Inc.\ under cooperative agreement with the National Science foundation.}

\subsection{Spectroscopy}\label{spectro}
The Huygens region of the Orion Nebula has been the subject of many spectroscopic studies. Most of these did not have velocity resolutions to determine with sufficient accuracy the radial velocities of features moving about the speed of sound (about 10 \kms) and the studies with this necessary spectral resolution usually treated only certain sections of the nebula. More complete high resolution mapping of the Huygens region have been made recently \citep{doi04,gar07,gar08} in the strongest lines. Those multi-slit-setting maps were made with primarily north-south oriented slits, so that the spatial resolution was excellent along the slit length, but poorer in right ascension because the spacing was usually wider than the characteristic ``seeing". In this study, we draw upon some of those earlier spectra, but our primary target was an extremely detailed east-west region centered on the Orion-S area and new observations were necessary.  

\begin{figure*}
\epsscale{0.7}
\plotone{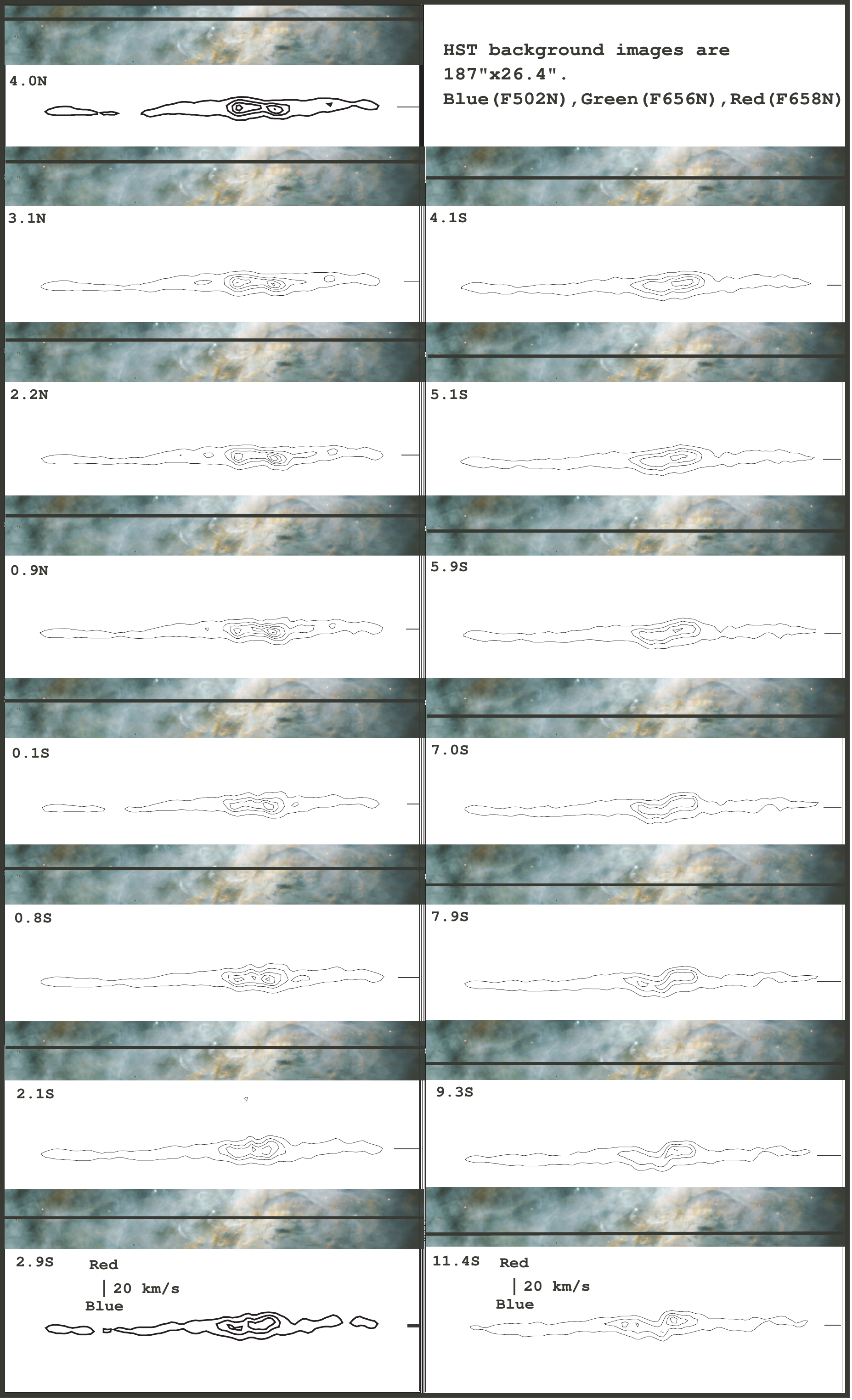}
\caption{
These187\arcsec x 26.4\arcsec\  samples of a WFPC2 image \citep{ode96} have superimposed the locations of the 15 slit spectra obtained. The declination shifts in arcseconds relative to 159-350 are shown for each setting.  Below each image is a contour plot of the strong (6583 \AA) [N~II] line beginning at 2000 counts and increasing in steps of 3000 counts. By setting the initial contour at 2000 counts one sees the velocity changes in the nebula more clearly, at the expense of the visibility of the faint high velocity features.  The short horizontal line in each contour image is the position for a heliocentric velocity of 20 \kms. Full resolution versions of all of the figures in this paper are included with the electronic edition of this journal article.
\label{fig:contoursNII}}
\end{figure*}

These new observations were made over the period 11--14 January 2007. The instrument configuration and performance was the same as in our earlier study of the Ring Nebula \citep{ode07}. The spectrograph was employed in a single spectral order mode by isolating the order containing the \Halpha\ 6563 \AA\ and [N~II] 6548 \AA\ and 6583 \AA\ lines with a narrow-band interference filter. The 70 $\mu$m wide entrance slit projected 0.9\arcsec\ onto the plane of the sky. We processed the well exposed sample of 187\arcsec. The 24 $\mu$m pixels of the  Site 1024x1024 CCD detector were double-sampled along the slit (0.62\arcsec\  per pixel). The  spectrum was read out in single pixels along the dispersion direction, giving Full Width at Half Maximum (FWHM) of the Thorium-Argon comparison lamp emission lines of about 7.0 \kms.  The atmospheric seeing varied from one to one and one half arcseconds during the exposures. A series of east-west oriented slit settings were made using offsets from the bright star 159-350=JW 499 (in the position based system introduced in \citet {ode94} or the catalog of \citet{jw85}) and at intervals of about 1\arcsec\ at locations shown in Figure \ref{fig:contoursNII}. We will use the position based system of designation throughout this paper, adding decimal places when an accuracy of greater than about 1\arcsec\ is needed. Exposures of 600 s were made with characteristic peak signals of 30,000 counts in \Halpha. Transparency variations were corrected by normalizing each slit's signal against a similar region of a flux calibrated \citep{ode99} image of this object \citep{ode96} made with the HST's WFPC2, but gaussian blurred to resemble ground-based image quality. Figure \ref{fig:contoursNII} shows contour plots of the resulting spectra. 

\begin{figure*}\centering
\includegraphics[width=\textwidth]{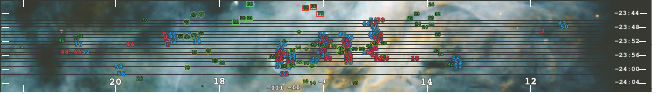}
\caption{
This is the same WFPC2 image as Figure1 but showing all of the slit settings. The labels on the short axis are the Declination values south of 5\arcdeg\ while the labels on the long axis are the Right Ascension values east of 5$\rm ^{h}$32$\rm ^{m}$ (both epoch 2000).  The -I, -II, and -III refer to features within the multiple eastward moving shocks that form HH 529.
The letters J show the locations of spectral features that must be microjets since they show a range of velocities starting at the system velocity, blue indicates \Halpha\ and red [N~II].  The numbers indicate the heliocentric radial velocities and all of them are negative, that is, the minus signs have are not shown.
The green numbers indicate high velocity features in \Halpha\ and/or [OIII] found in \citet{doi04}. The other numbers show results from our slit spectra.
\label{fig:slitResults}}
\end{figure*}

Because of the high velocity resolution and excellent signal level, it was possible to search for faint
high velocity features. The results of this search are shown in Figure \ref{fig:slitResults}, which superimposes the velocities of high velocity components upon the slit settings and a WFPC2 image. For completeness, this data also shows the results from an earlier study \citep{doi02}.

\subsection{Imaging}\label{imaging}
Our imaging program was planned to allow a homogenous set of data to be obtained, which was intended to allow the best possible comparison of earlier and later images. In planning the observations it was recognized that the high spatial resolution of the ACS-WFC (0.05\arcsec / pixel in the drizzled images) was better than that of the WFPC2 (0.0996\arcsec / pixel), but the latter instrument allows a longer time base since it was installed earlier. The planning became more complex as the ACS-WFC failed and WFPC2 observations were substituted where possible. 

\subsubsection{New Observations}\label{newimages}
\begin{deluxetable*}{ccccccc}
\tabletypesize{\scriptsize}
\tablecaption{Program GO 10921 Data Used in this Study \label{tab:newdata}}
\tablewidth{0pt}
\tablehead{
\colhead{Visit} & \colhead{Target} & \colhead{Camera} & \colhead{Date} & \colhead{Modified Julian Date} & \colhead{Filter} & \colhead{Exposure (seconds)}}
\startdata
1 & Orion-S & WFPC2 & 2007-12-25 & 54459.3 & F502N & 300\\
-  &        -       &     -          &             -         &       -        &  F547M & 90\\
-  &        -       &     -          &             -         &       -        &  F631N & 600\\
-  &        -       &     -          &             -         &       -        &  F656N & 180\\
-  &        -       &     -          &             -         &       -        &  F658N & 300\\
-  &        -       &     -          &             -         &       -        &  F673N & 480\\
2 & HH 528 & WFPC2 & 2007-09-16 & 54359.8 & F502N & 400\\
-  &        -       &     -          &             -         &       -        & F547M & 120\\
-  &        -       &     -          &             -         &       -        & F631N & 600\\
-  &        -       &     -          &             -         &       -        & F656N & 240\\
-  &        -       &     -          &             -         &       -        & F658N & 400\\
-  &        -       &     -          &             -         &       -        & F673N & 540\\
5 &   HH 502& ACS-WFC& 2006-11-22 & 54061.4 & F502N & 1300\\
-  &        -       &     -          &             -         &       -        & F658N & 945\\
-  &        -       &     -          &             -         &       -        & F660N & 1000\\
6 &   HH 540 & WFPC2&   2007-11-05 & 54409.7 & F502N & 1000\\
-  &        -       &     -          &             -         &       -        & F547M & 80\\
-  &        -       &     -          &             -         &       -        & F656N & 600\\
-  &        -       &     -          &             -         &       -        & F658N & 800\\
-  &        -       &     -          &             -         &       -        & F673N & 800\\ 
\enddata
\end{deluxetable*}

In this paper we report on the analysis of new images in four areas of the Orion Nebula, supplemented
by similar investigation of pairs of earlier observations. The four areas were selected to coincide as much as possible with earlier images of what were thought to be the most interesting and relevant objects. The characteristics of the new observations are summarized in Table~\ref{tab:newdata}. The total exposure times presented were obtained in triple exposures, which allowed removal of cosmic ray induced spots on the images. On-the-fly routinely processed data were the starting point for our study.

\subsubsection{Earlier Observations}\label{oldimages}
\begin{deluxetable*}{ccccccc}
\tabletypesize{\scriptsize}
\tablecaption{Archived Observations Used in this Study \label{tab:olddata}}
\tablewidth{0pt}
\tablehead{
\colhead{Program} & \colhead{Target} & \colhead{Camera} & \colhead{Date} & \colhead{Modified Julian Date} & \colhead{Filter} & \colhead{Exposure (seconds)}}
\startdata
GO 5193 & Orion-Southeast & WFPC2 & 1993-12-29 &49350.8 & F502N & 800\\
-  &        -       &     -          &             -         &       -        &  F547M & 200\\
-  &        -       &     -          &             -         &       -        &  F656N & 700\\
-  &        -       &     -          &             -         &       -        &  F658N & 1200\\
GO 5469 & HST1,HST10,LV3 & WFPC2 &1995-03-21 & 49797.5 & F502N & 300\\
-  &        -       &     -          &             -         &       -        & F547M & 90\\
-  &        -       &     -          &             -         &       -        & F631N & 300\\
-  &        -       &     -          &             -         &       -        & F656N & 180\\
-  &        -       &     -          &             -         &       -        & F658N & 300\\
-  &        -       &     -          &             -         &       -        & F673N & 300\\
GO 8121 &   HH 203-204 & WFPC2&   2000-03-20 & 51633.1 & F502N & 780\\
-  &        -       &     -          &             -         &       -        & F656N & 780\\
-  &        -       &     -          &             -         &       -        & F658N & 1800\\
-  &        -       &     -          &             -         &       -        & F673N & 1800\\ 
GO 8121 &   BN-KL & WFPC2&   2000-09-13 & 51800.9 & F502N & 780\\
-  &        -       &     -          &             -         &       -        & F656N & 780\\
-  &        -       &     -          &             -         &       -        & F658N & 1800\\
-  &        -       &     -          &             -         &       -        & F673N & 1800\\ 
GO 9460 &   HH 502& ACS-WFC& 2002-08-07 & 52493.7 & F502N & 2000\\
-  &        -       &     -          &             -         &       -        & F658N & 1000\\
-  &        -       &     -          &             -         &       -        & F660N & 1000\\
\enddata
\end{deluxetable*}

We drew on a large body of earlier observations made with the WFPC2 and the ACS-WFC. Their characteristics are presented in Table~\ref{tab:olddata}. This listing includes not only the images matched with the new observations, but also images that were reprocessed and analyzed as part of this program. In addition to these long time-base ACS-WFC images, we also employed archived F658N ACS-WFC images obtained through program GO 10246 with Massimo Robberto as principal investigator.

\subsubsection{Processing the Image Data}\label{processing}  
The first step in comparison of the first and second epoch images was to align the two. In each case we took the first epoch images as the standard and shifted the second epoch images. Since the individual CCD detectors of both cameras can change with respect to their neighbors, alignment and comparison was only made using reference stars falling into a single CCD in both  the early and later images. This is the same technique used in our earlier investigations \citep{bal00,doi02,od03} using data with a shorter time interval. The accuracy of the alignment was typically about 0.2 pixels, which corresponds to about the 3 \kms\ internal dispersion of velocities of the Orion Nebula Cluster stars \citep{jw85,fur08}. Because the first and second epoch fields of view were not exactly on the same part of the sky, the region useful for comparisons is more irregular and smaller than the second epoch images.

After alignment the signal of the two images was normalized and a search was made for moving objects by taking a ratio of the first and second epoch images. In the case of identical images, the ratio image would be a smooth value of unity and in the case where objects are moving there would be a pattern with a dark leading edge and a bright trailing edge. In most cases motion of individual features was measured using the least squares method \citep{har01} employed previously,  or when the objects were large or the regions complex, simply measuring their positions directly. The measurement uncertainties ranged from 0.1 to 0.4 pixels, depending on the nature of the object being measured. The uncertainty in the direction of the motion is less well defined, but can be 30\arcdeg\ for very small motions and only a few degrees for the large motions. The results of the measurement of discrete objects are presented in Table~\ref{tab:huygens} (Appendix B) and Table~\ref{tab:hh502} (Appendix C). However, there were many large-scale moving features not measured in either fashion, but the reality of the motion in the ratio images is unquestionable. 

\begin{figure*}
\epsscale{1.0}
\plotone{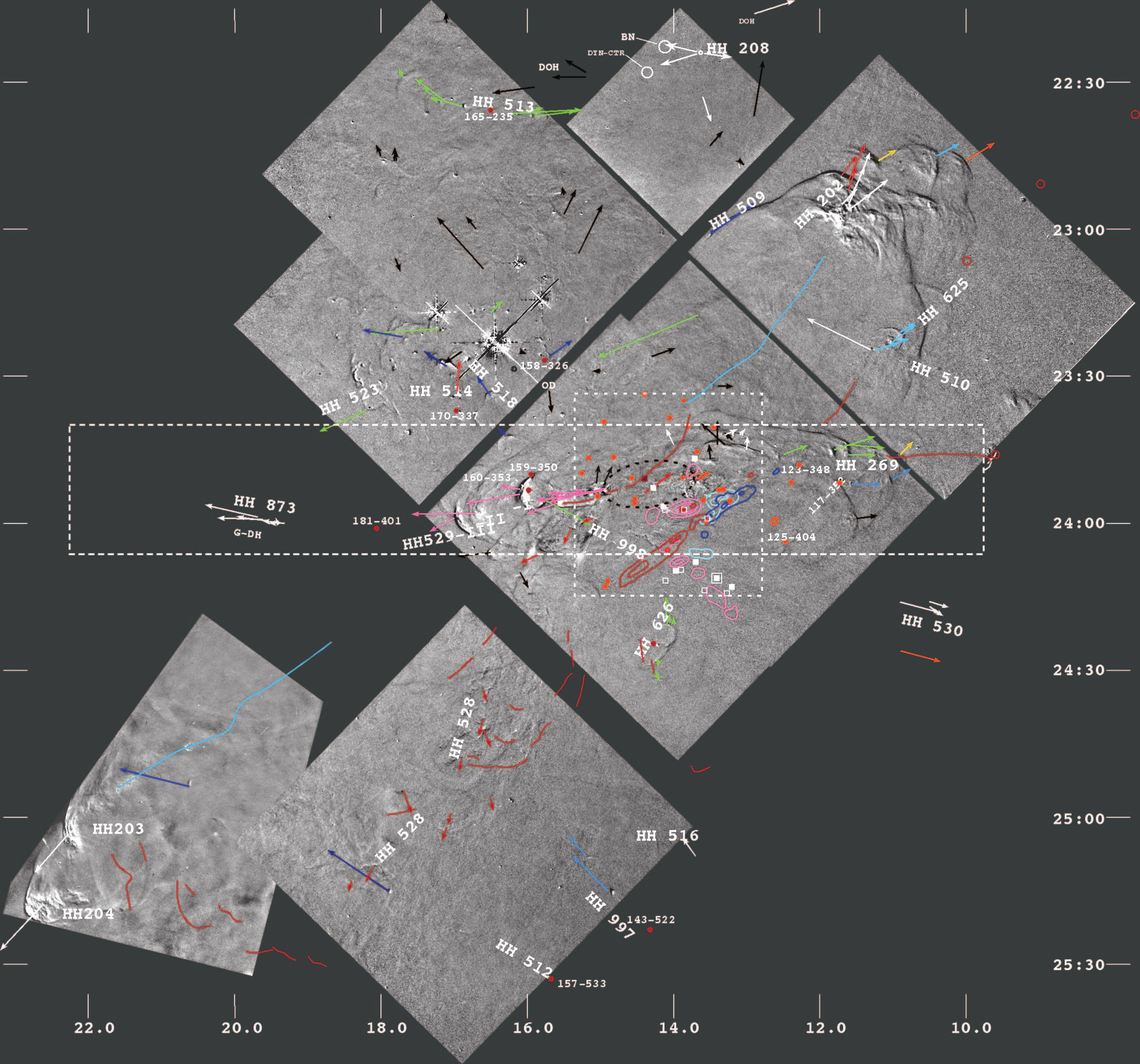}
\caption{
The determination of tangential motions for the Huygens region are shown in this 213\arcsec x219\arcsec\ mosaic of F656N ratio images (first/second), with north up. See the text for details of the labeling. The nearly square central dashed region is shown enlarged in Figure 7 and the elongated dashed region is the spectroscopy field in Figure 1 and Figure 2.
 \label{fig:master656}}
\end{figure*}

\begin{figure*}
\epsscale{1.0}
\plotone{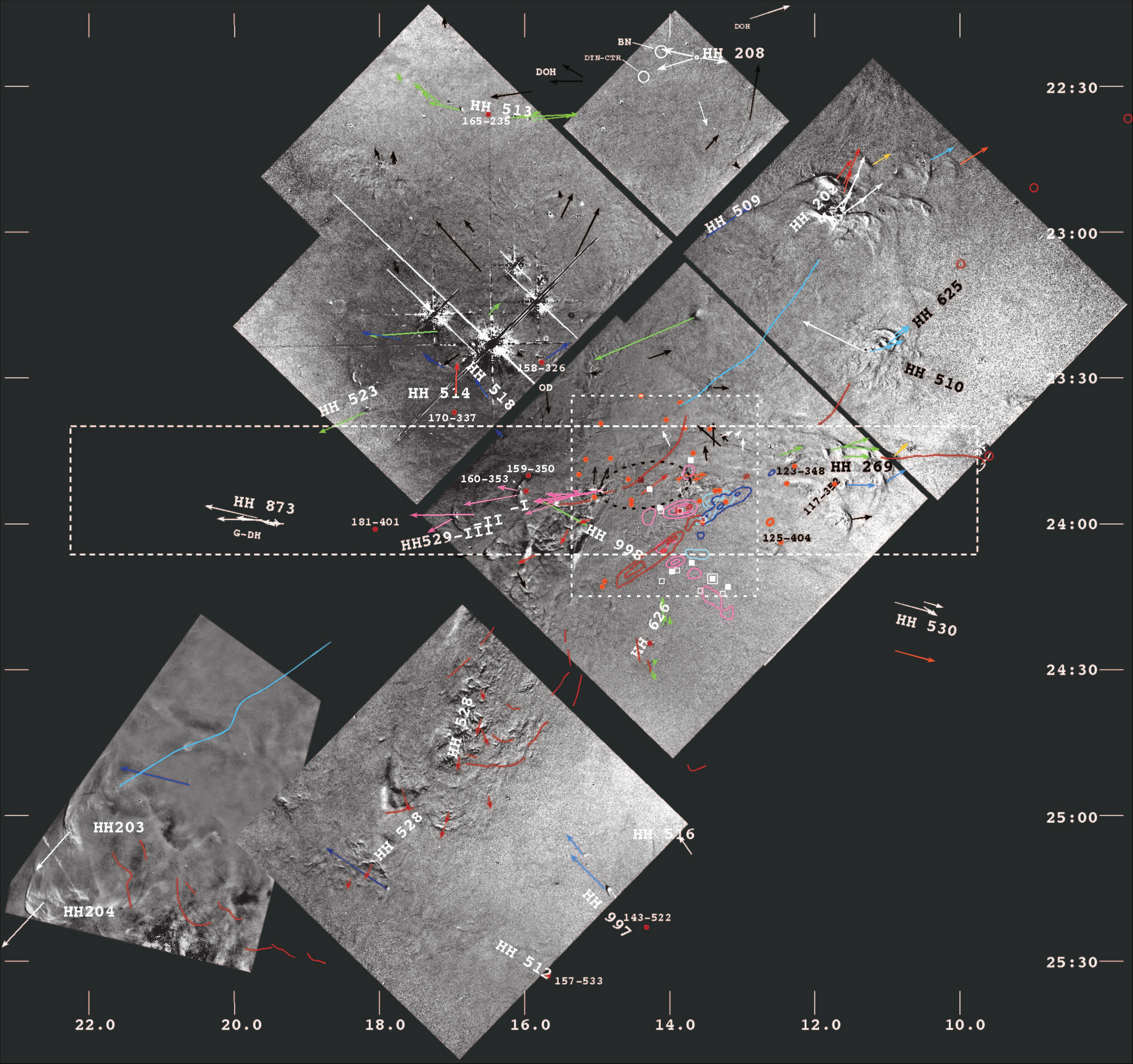}
\caption{
The same as for Figure 3, except this shows the F658N ratio images.
\label{fig:master658}}
\end{figure*}

\begin{figure*}
\epsscale{1.0}
\plotone{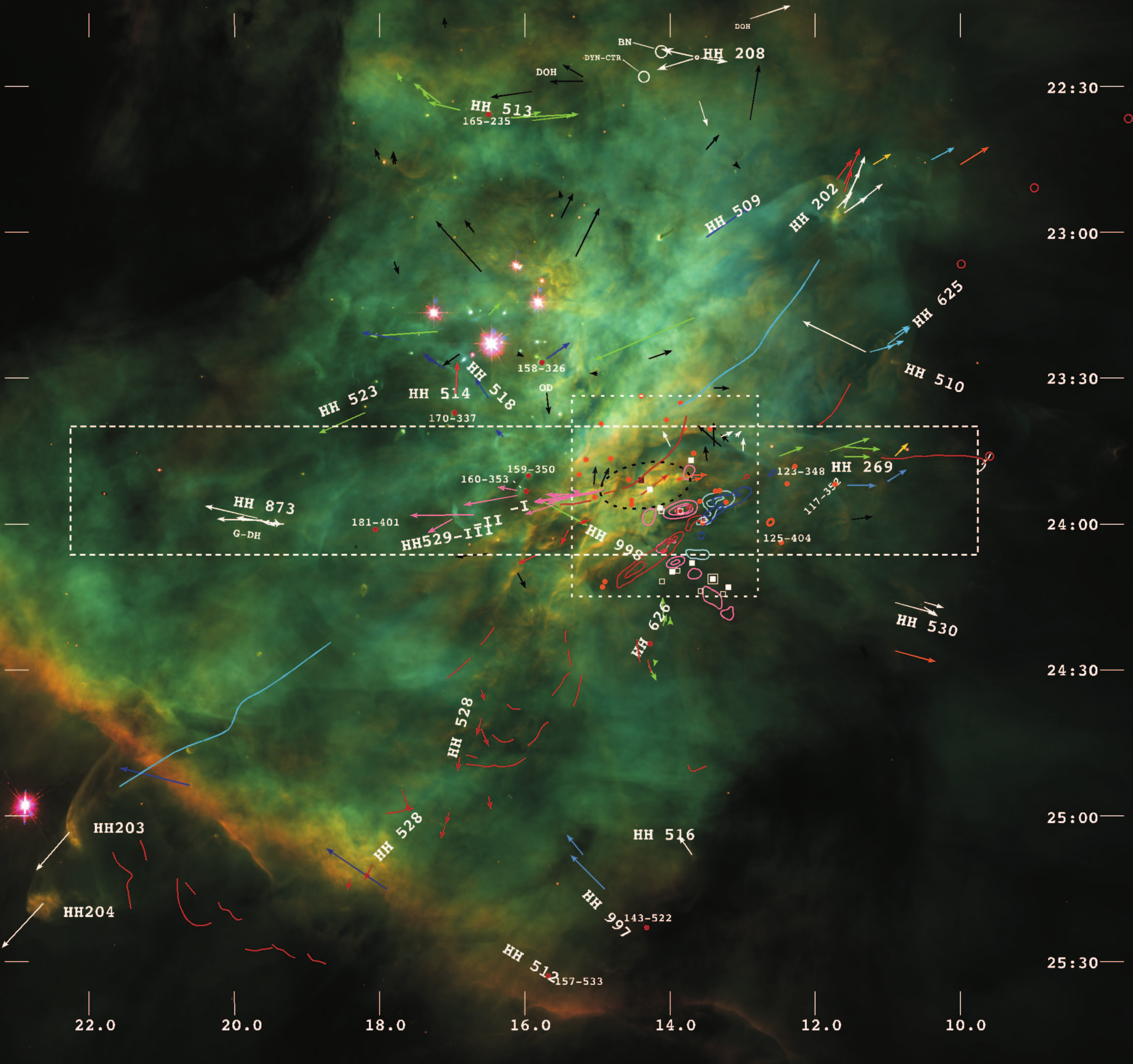}
\caption{
The same as for Figure 3, except the background image is a section of the WFPC2 mosaic image \citep{ode96}. 
\label{fig:masterColor}}
\end{figure*}

Figure 3 and Figure 4 show ratio images for our Visit 1 and Visit 2 images for the WFPC2 F656N and F658N filters, supplemented by images near HH 203 and HH 204 from a comparison of  archived GO 5193 and GO 8121 images which have been smoothed by median filtering. The difference in time between the first and second epoch images was 12.76 years for Visit 1, 12.49 years for the Visit 2 F631 and F673N images, 13.71 years for the Visit 2 F502N, F547M, F656N, and F658N  images, and  6.25 years for the HH 203/204 images.  At our adopted distance of 440 pc (c.f. Appendix A), these intervals correspond to WFPC2 velocity scales of  16.3, 16.6, 15.2, and 33.2 \kms\ per WFPC2 pixel respectively.  Ratio images in F502N were created and used for finding rapidly moving objects but are not shown here as all moving features in this filter are also seen in the F656N image, whereas many of the F656N moving features are not seen in the F502N image. Figure 5 shows the same tangential velocity image superimposed on a WFPC2 color image.

The detailed coding of the superimposed data for Figure 3, Figure 4, and Figure 5 are described here, rather than in the figure captions. The markings on the outer borders show the Declination south of \(-5\arcdeg\) and the Right Ascension east of 5$^{\rm h}$ 35$^{\rm m}$. Filled circles indicate the positions of stars mentioned in the discussion and those within the dashed outline are found in the near infrared catalog of \citet{hil00}.  Arrows indicate the direction and magnitude of the tangential velocities, with the arrow length corresponding to the motion in 300 years. The colors of the arrows are added only to indicate local groupings, the same color being used multiple times for different groups. The dashed elongated rectangular outline indicates the region of radial velocity results shown in Figure 2. Open red circles indicate the position of \htwo\ knots \citep{sta02} and red lines \htwo\ features seen in Subaru images \citep{kai00}. The long irregular blue lines delineate collimated high velocity flow as indicated by He~I 10830 \AA\ emission \citep{tak02,hen07}.  The strong dark red/blue contoured lines indicate CO outflow \citep{zap05} and the pastel colored pink/light-blue contoured lines indicate SiO outflow \citep{zap06}.  Open squares indicate the positions of H$_{2}$O maser sources \citep{gau98}. Point sources within the dashed outline are coded by the shortest wavelength of their detection, with  filled white squares indicating the positions of radio only visible sources \citep{zap04a,zap04b,zap05}, red squares the positions of sources seen only in the mid infrared \citep{smi04,rob05}, and filled orange circles positions of stars in the near infrared catalog of \citet{hil00}.  A few motions outside of our field of view have been added with the coding OD=\citep{od03}, DOH=\citep{doi02}, and G-DH=\citep{gar07}. The region most likely to contain the sources of the high velocity optical outflows is shown as a dark dashed ellipse.

\begin{figure*}
\epsscale{1.0}
\plotone{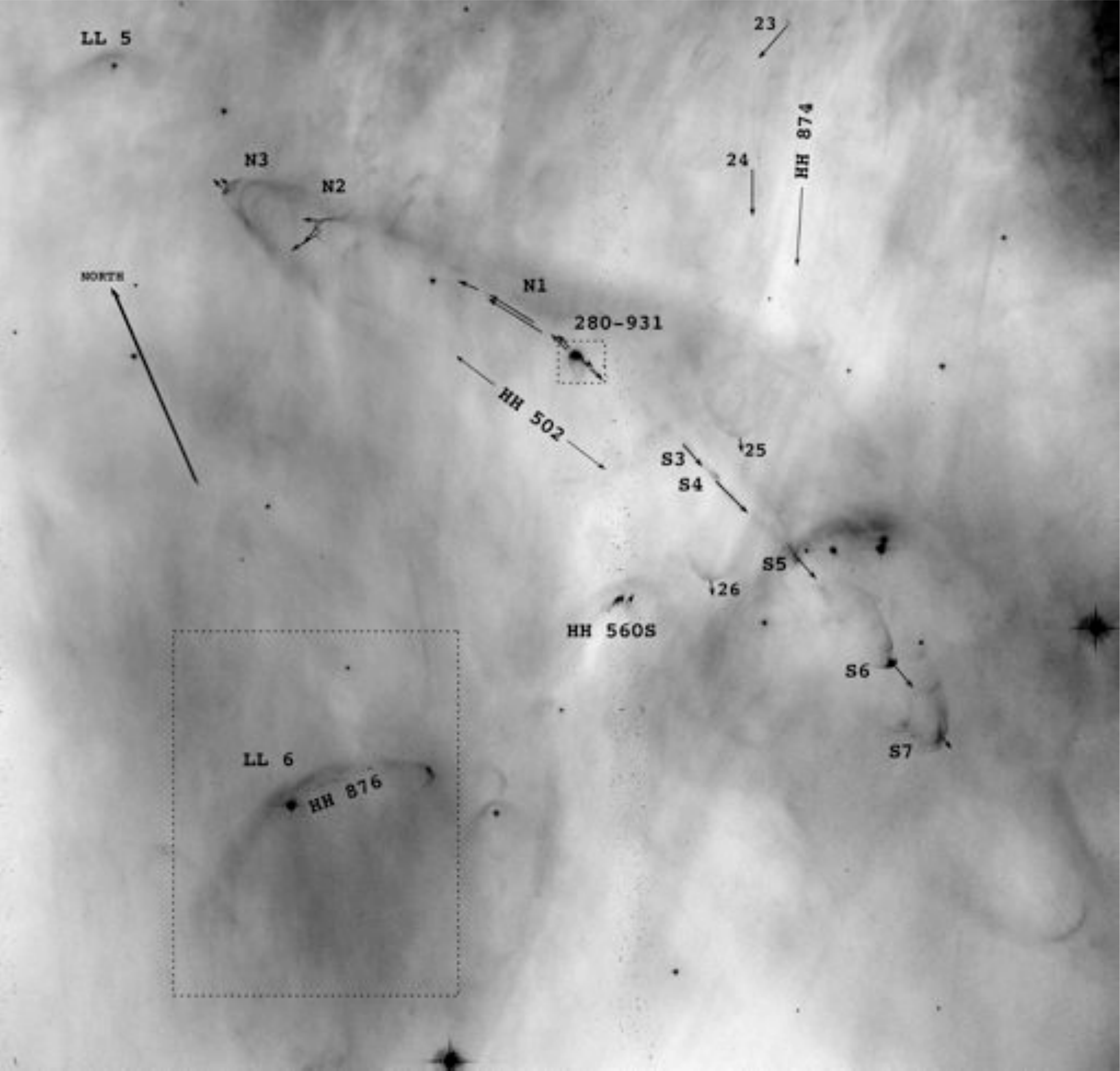}
\caption{
This 187\arcsec x178\arcsec\ inverted brightness ACS-WFC F658N image of the region around HH 502 has PA=-23.4\arcdeg\ as up. Motions over a period of 100 years are shown. The dashed boxes show the regions depicted in larger scale in Figure 10 and Figure 11. The vertical strip with many small dark features is a zone of poor removal of cosmic ray events.
\label{HH502wide}}
\end{figure*}

Figure 6 shows the Visit 5  region of overlap with the GO 9460 field near HH 502. In this case the time interval was 4.29 years and the velocity scale 24.3 \kms\ per ACS-WFC pixel.  It has had the faint fine features enhanced by division by a third order polynomial fit to the background. The nomenclature is that of \citet{bal06}. 

\section{Discussion}

This rich body of new data allows us to draw conclusions about many aspects of gas-flow and star formation in the Orion Nebula Cluster (ONC) and its secondary star formation regions in the BN-KL and Orion-S regions. In this discussion we will first consider outflows associated with the Orion-S region (both as individual systems and as a possible global unit). We then treat what the new data tell us about the BN-KL region. The multiple outflows to the south of the Huygens region and centered on HH 502 are discussed third. The discussion concludes with consideration of individual proplyds and outflow systems. Throughout this paper we will assume that the velocity of the host Orion Molecular Cloud (OMC) is \Vsun=25.8 \kms\ \citep{ode08} and that of the ONC is 25.6 \kms\  \citep{fur08}.  Velocities relative to the OMC will be designated as \Vomc. We will continue to assume that the distance is 440 pc, so that one-tenth parsec subtends 46.9\arcsec. Position Angles (PA) will be expressed in the ordinary way, counter-clockwise from north and $\theta$, the angle of flow with respect to the plane of the sky, is positive in the direction of the observer. We draw heavily on published slit-spectroscopy maps of higher ionization \citep{doi04} and lower ionization lines \citet{gar07} (henceforth DOH04 and GD07), and where we do not have new tangential velocities, on \citet{doi02}, henceforth DOH, or \citet{od03}, henceforth OD03.

\subsection{Outflow from the Orion-S Region Near-Surface Sources}
There is by now a rich literature about large-scale (greater than 0.1 pc) outflows probably originating from the Orion-S region, the idea going back to the symmetric forms of HH 202 and HH 203+HH 204 being pointed in opposite directions across Orion-S, then developing as more outflows (HH 269, HH 529, HH 528, HH 625) were found. \citet{hen07}, henceforth HO07, have most recently summarized the history and  characteristics of these outflows, including their possible connection to much larger apparent shocks outside of the Huygens region, so that it is not necessary to discuss each object  anew, beyond summarizing how the new observations have altered our understanding. Our new observations do give valuable new information about the exact region of the sources of these outflows. 

\begin{deluxetable*}{lccrccccr}
\tabletypesize{\scriptsize}
\tablecaption{Motions of Large-Scale HH Objects Originating near Orion-S
  \label{tab:orion-S}}
\tablewidth{0pt}
\tablehead{
\colhead{Designation} & \colhead{\Vt}  & \colhead{source} & \colhead{\Vsun} & \colhead{source} &\colhead{Vector Velocity} & \colhead{$\theta$} & \colhead{PA (degrees)} & \colhead{Timescale (years)}}
\startdata 
HH 202-South &  59  &  GO 10921   &   \(-38\) & DOH04 &  89 & 48 & 326 & 2,600\\
HH 202-North &   45  &  GO 10921    &   \(-21\) & DOH04 &  67 &  47& 336 & 3,600 \\
HH 202-NW    &   31  &  GO 10921    &   \(-18\)  & DOH04 &  55 & 56 & 310 & 5,600\\
HH 202-WNW &  38  &  GO 10921   &   \(-20\)  & DOH04 &  61 & 52 & 298 & 5,100\\
HH 203             &  77  &   DOH02       &   \(-46\)  & DOH04 & 77  & 43 & 139 & 3,400\\
HH 204             &  92   &  DOH02      &    \(-18\)  & DOH04 & 92  & 26 & 137 & 3,400\\
HH 269              &  50  &  GO 10921  &   \(-46\)  & GO 10921 & 88 & 55 &  275 & 1,800\\
HH 269-Ram   &   18  & GO 10921  &    \(-25\)   & GO 10921 & 54 & 71 & 290 & 1,500\\ 
HH 528-Base   &   22  & GO 10921 &    17     & HO07        &  24 & 22 & 178 & 5,700\\
HH 528-Cap    &    24  & GO 10921 &    17     & HO07        &  24 & 22 & 159 & 5,500\\
HH 529-III         &    40  & GO 10921 &   \(-30\)    & GO 10921 & 69  & 54 & 119 & 2,000\\ 
HH 529-II          &    92  & GO 10921 &   \(-31\)    & GO 10921 & 108 & 32 & 91  & 780\\
HH 529-I           &    54  & GO 10921 &   \(-30\)   & GO 10921  &  78   &  46 & 90 & 980\\
HH 529-Ram    &    59  & GO 10921 &  \(-44\)   & GO10921   &   84   & 45  & 95 & 410\\
\enddata
\tablecomments{All velocities are in \kms\ and Timescales are calculated from the tangential motions as if the point of origin was at 144-352.  GO 10921 indicates data from this study.}
\end{deluxetable*}

Once again we find the well known result that all of the large-scale outflows are blue-shifted with respect to the OMC. When their current motions are projected backwards, these projected lines cross in the Orion-S regions, suggesting that they have a common origin in one or more objects located there. The results for these objects are summarized in Table~\ref{tab:orion-S}. The region that probably contains the sources of the high velocity optical outflows is shown as an ellipse in Figure 3, Figure 4, and Figure 5. This region is a more refined estimate of the location and properties of the OOS (Optical Outflow Source) identified in \citet{od03}.

\paragraph{HH 202.} HH 202 is at the apex of a large parabolic high ionization feature, as shown in Figure 5.  The variety of ionization stages indicates that although it forms as a collimated high velocity outflow it also contains neutral material. This can be due to the flow impacting neutral material or that the flow has compressed the ambient ionized gas to the degree that it traps the ionization front. If the latter, this would require that the material lies within the main cavity of the nebula rather than in the foreground veil.  The white arrows in this region in Figure 5 indicate tangential motions in the HH 202-South brightest part of the object and the red arrows the HH 202-North part. The yellow arrow indicates flow in a shock feature ahead of HH 202-North (designated as HH 202-NW) and the blue and orange arrows two shocks even further ahead (designated as HH 202-WNW).  The collimated flow probably driving these shocks has been measured by DOH04 to have radial velocities of about  \(\Vomc =-40~\kms\)  to \(-60~\kms\) and the details of this flow are best shown in the He~I 10830 line \citep{tak02}, which is seen in good contrast against the background emission from the nebula because the line is usually quite optically thick except where the emitting materially is Doppler shifted off the core of the line. 

\paragraph{HH 203.} HH 203 is a well defined bow shock with low ionization features at its tip. It is apparently driven by a high velocity, high ionization jet that emerges into the zone ionized by \ori\  or the nearer (in the plane of the sky) \oriA\ (which dominates the ionization of some of the proplyds in this area) about half the distance from the point of origin.  The jet radial velocity is \(\Vsun =-28~\kms\).

\paragraph{HH 204.} HH 204 is nearly at the same PA as HH 203, but shows marked differences.
Its apex is a flocculent structure that includes low ionization features even though the body of the enclosed parabolic form has extended [O~III] emission, indicating that it is forming in ionized gas. No driving jet is seen. It should be noted that HO07 found a broader shock front extending beyond even HH 204 and at a PA midway between HH 204 and the terminus of HH 528, as if there had been three events of collimated flow in about this direction. 

\paragraph{HH 269.} HH 269 is composed of a series of well defined, wide-ionization range, shocks oriented almost due west \citep{wal95}.  In this study, only features near 116-345 were measured (the so-called HH 269-East shock) but both in tangential and radial velocity. A 5\arcsec\ narrow but wiggly low ionization feature was also measured  at its western terminus and is designated here as HH 269-Ram.

\paragraph{HH 528.} HH 528 shows a broad band of irregular, low ionization,  structures headed southeast from the Orion-S region.  It is composed of two regions, the base or ``jet'' (as designated in HO07) and the southeast-most bow shock (again as designated in HO07).  The term ``jet'' only loosely applies as this feature is much broader than any collimated flow producing the other HH objects in this study and one sees individual bow shocks within the object.  Falling back on the original discovery work notation for HH 528 \citep{bal00} where it was noted that the entire object is shaped like a mushroom, we will here designate the ``jet'' as the ``base'' and the end feature as the ``cap''.  HH 528 may in fact represent two outflows, the ``base'' object  defined by broad irregular structure, oriented towards PA=155\arcdeg\ and containing features moving towards PA=178\arcdeg\ and a second outflow represented by the ``cap'', oriented very approximately towards 147\arcdeg\  and containing features moving towards 159\arcdeg.  In both cases the objects  may be passing through the low-ionization portion of the nebula, rather than beomg within the main body of photoionized gas. A mediating argument against division of HH 528 is that the velocities and spatial orientation of the two components are very similar.  

There is a peculiar complex of features around 5:35:17.6 \(-5\):24:54 (all coordinates in this paper will be in epoch 2000). Parts of this complex seem related to the HH 528 flow, such as the 177-454 shock, which is measured in [S II] to be moving at 37 \kms\ towards PA=199\arcdeg. However, there are other features that seem to be moving at velocities of order 130 \kms\ towards PA = 100\arcdeg. Most puzzling is a feature just to the southeast of 177-454,  where a naive interpretation of the [N~II] ratio map would imply
an oppositely directed motion of 130 \kms\ towards PA = 280\arcdeg. However, careful inspection of the individual images in multiple lines suggests that this feature is also moving towards PA= 100\arcdeg. The reason for the discrepant behavior in [N~II] is that in this filter the feature seems to be a moving dark feature rather than a moving bright feature, which reverses the sense of motion in the ratio image. In addition, the morphology of the feature seems to be evolving, becoming darker in the second epoch. It is much larger and more diffuse than the shock features associated with HH 528 and at this point remains unexplained.

\paragraph{HH 529.} HH 529 is composed of a series of shocks oriented towards the east  and moving in that direction.  Here we break it down into designated shocks HH 529-III, HH 529-II, and HH 529-I proceeding from east to west (as shown in Figure 2--Figure 5). There is a series of small, rapidly moving features which may represent the driving, nearly collimated outflow that we designate as HH 529-Ram.  There may be an HH 529-0 shock which is not obvious on images, but reveals itself through high radial velocity features a few arcseconds north and south of the HH 529-Ram group (cf. Figure 2).  The HH 529-Ram features begin at a Right Ascension of 5:35:14.91 and end at 5:35:15.6 while the putative HH 529-0 shock is at 5:35:15.5, i.e. near the west end of the Ram features, and its velocity \(\Vt = -48~\kms\) is comparable to the Rams' \(\Vt =-44~\kms\), which also argues for an association. Bow shocks HH 529-III, HH 529-II, and HH 529-I are high ionization only, indicating that they fall within the zone of ionization by \ori, a conclusion supported by the detailed spectroscopic analysis of HH 529-III+HH 529-II by \citep{bla06}.  However, the HH 529-Ram feature is seen in both high and low ionization lines, indicating that this outflow is moving from a low into a high ionization region.

\subsection{Outflow from the Orion-S Deeply Imbedded Sources}
\subsubsection{Molecular Outflows}
The brightest infrared and radio sources lie to the southwest of the putative source or sources producing the multiple large-scale HH objects associated with Orion-S. Recent radio observations reveal two well defined molecular outflows in this southwest part of Orion-S. CO observations \citep{zap05} detail a bipolar outflow with the blueshifted component towards PA=305\arcdeg\ and apparently arising from the infrared star 136-400, which has associated H$_{2}$O masers \citep{gau98}.  A second study in SiO reveals a bipolar molecular outflow that apparently arises from the infrared star 136-355, with its blueshifted component towards PA=284\arcdeg\ \citep{zap06}.  These features are shown in Figures 3--5. There are several spots of SiO emission in the strong radio sources 137-408 (CS 3) and 134-411 (FIR 4), but no obvious directed outflow. 136-400 and 136-355 have no optical counterparts and they must lie behind a significant amount of obscuring material, so one would expect to find optical components to the molecular outflows associated  with only the blueshifted components.

The SiO blueshifted component coming from 136-355 intersects with the HH 269 series of shocks, which have an orientation of 273\arcdeg\ \citep{wal95} , an orientation strengthened by the presence of long \htwo\ filaments \citep{kai00}. The feature for which we have measured the tangential velocity lies upon a projection of the SiO blueshifted outflow, but the direction of motion of these features (PA=275\arcdeg) more nearly agrees with the overall orientation of HH 269 and a line projected back to the HH 269-Ram feature.

\subsubsection{The Optical HH 625 Object Associated with the High Velocity CO Outflow}
There is a well defined series of optical and infrared features associated with the CO blueshifted component. One sees \htwo\ features extending towards 136-400 and there are a series of \htwo\ knots \citep{sta02} extending out to 204\arcsec\ (0.44 pc). There is a peculiar, low ionization, optical feature that has been designated as HH 625 and we measured four features in it with an average motion of \Vt =34 \kms\ towards PA=296\arcdeg. We have examined \Halpha\ and [N~II] spectra of this region \citep{doi04} and determined that \(\Vsun =-8~\kms\). This means that this optical feature (which must be formed in a region of lower obscuration than applies closer to 136-400) has a velocity vector of 48 \kms\ inclined 41\arcdeg\ towards the observer with respect to the plane of the sky. The CO observations trace the blueshifted component to \(\Vsun = -36~\kms\) (\(\Vomc = -62~\kms\)), which means that the optical feature probably represents mass-loaded material moving more slowly than the original outflow.  However, the small size of the \htwo\ knots indicates that the flow that drives them has remained very narrow and collimated.

\subsection{Outflows from the Central Region}
\begin{figure}
\plotone{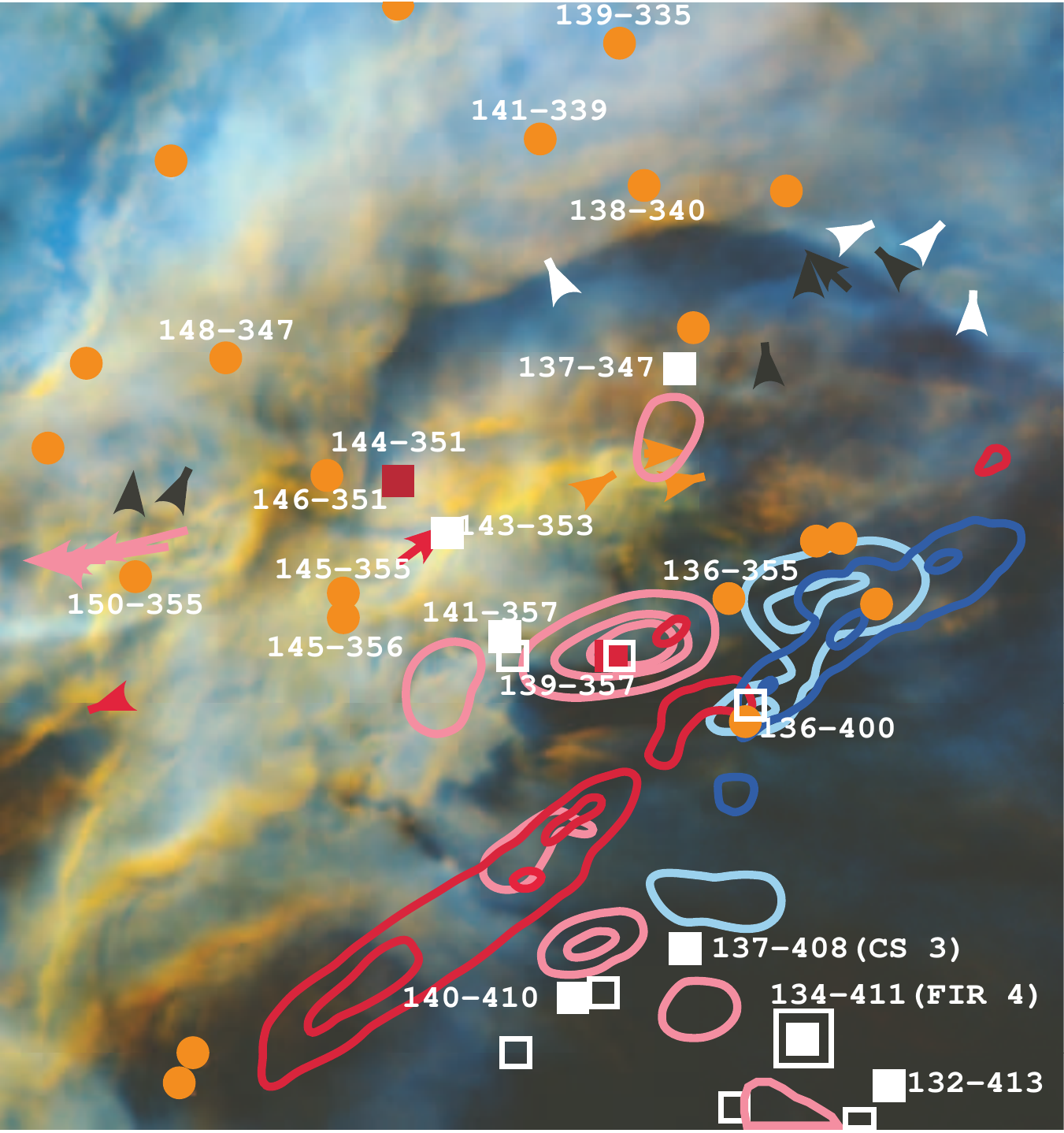}
\caption{
This 38.7\arcsec x41.0\arcsec\ sample of the central region shown in Figure 5 uses the same color coding (although a different color balance) and the identification numbers of the sources have been added. Motions projected over 100 years are now shown.  The symbols have the same meaning as in Figure 3.  
\label{CentralRegion}}
\end{figure}

In Figure 7 we show the central portion of the Huygens region that includes Orion-S. This incorporates all of the area where the axes of symmetry of the flows (ordered counter-clockwise) of HH 202,  HH 529, HH 203, HH 204, HH 528, HH 269, and HH 625 intersect, this general area having been previously identified as the most likely source for these outflows \citep{bal00,od03,hen07}.  This figure is similar to Figure 5, except the color balance is slightly different, we have added labels identifying the individual sources, and tangential velocities are shown projected for only 100 years. 

Although we have derived higher accuracy tangential velocities, these have not narrowed down the identification of the sources of the large outflows except that various outflows seem to originate in the eastern part of this small field.  If the features we call in this paper the HH 269-Ram and HH 529-Ram are from the same source, its position is narrowed down to a Right Ascension interval of only 15\arcsec. However, to link these two flows as a single bipolar flow demands that one of the sides has been deflected towards the observer by about 116\arcdeg.  Although a mechanism that provides deflection as a beam passes through a density gradient has been proposed \citep{can96}, it seems a stretch to use this as the process that allows linking HH 269 and HH 529 into a single bipolar flow. The infrared source 146-351 and far infrared source 144-351 are the top candidates for the sources of these outflows.   Although the other outflows must also arise in this same region, linking them all into a single source requires invoking the same large amount of deflection multiple times, which is unlikely.  Multiple infrared and radio sources exist in this region and if they individually produce bipolar flows, then the observational selection effect of seeing only blueshifted outflow from a source lying behind the main ionization front would apply and explain the dominance of blueshifts.  If the extinction that causes the sources to be visible only in the infrared was local to the source, then one would expect to see both redshifted and blueshifted extended flow.

A recent infrared polarization study \citep{has07} may provide additional useful information. \citet{has07} argue that imbedded stars with bipolar outflows create dumbbell shaped cavities along the inner parts of the outflow and that starlight scattered from the surface of these cavities would have a characteristic polarization pattern. They have studied the infrared polarization in this region and find polarization signatures around 136-355, 145-356, and 144-351, making them prime candidates for outflows.
Possible outflow from 145-356 or the star 145-355 only 0.9\arcsec\ north of it is discussed in \S\ 3.8. The direction of the putative flow from these two stars is indistinguishably the same as the high velocity flow feeding HH 203.  The east side of the polarization pattern associated with 144-351 is generally oriented to be like a flow in the direction of the HH 529-Ram features, thus strengthening the argument that this is the source for the HH 529 flow.  

In the southwest portion  of Figure 7 one finds the sources of the molecular outflows, 136-400 producing the bipolar CO flow, whose northwest blueshifted portion creates the optical object HH 625 plus the series of H$_{2}$ knots ahead of it, and 136-355, producing the SiO outflow, portions of which may be in the features we are calling the western end of HH 269. The polarization pattern for 136-355 looks perpendicular to the SiO outflow direction, so its interpretation is not clear.  Neither of the brighter radio sources (137-408, CS 3; 134-411, FIR 4) in the corner of Figure 7 have clearly associated optical outflows, although the shocks constituting HH 530 and shown in Figures 3--5, appear to move away from the region of 137-408 and 134-411.  

\subsection{Outflow from the BN-KL Region}
Study of the BN-KL region is primarily the subject of infrared and radio investigations since most of outflowing material and all of the luminous central sources are invisible at optical wavelength. However, optical studies have aided in narrowing down the origin \citep{jw88, doi02}.  Study of the motion of the three most massive sources in BN-KL shows that they are expanding away from a common point \citep{rod05,gom05} which we label as ``DYN-CTR'' in our figures. The timescale for expansion of the optically visible shocks is 1000 years \citep{doi02}, although this may be shorter if the material has been decelerated \citep{lee00}. \citet{gom05} find a dynamic timescale of about 500 years.  The cause of this explosive decay and the physics relating it to the molecular outflow are still unresolved, but it is reasonable to link the outflow to whatever occurred there. An alternative view is that of \citet{beu08}, who argue that SMA1 is the source of the high velocity flow, with a dynamic timescale of about 1000 years, and that source I is the source of the low velocity outflow.

\begin{figure}
\epsscale{1.0}
\plotone{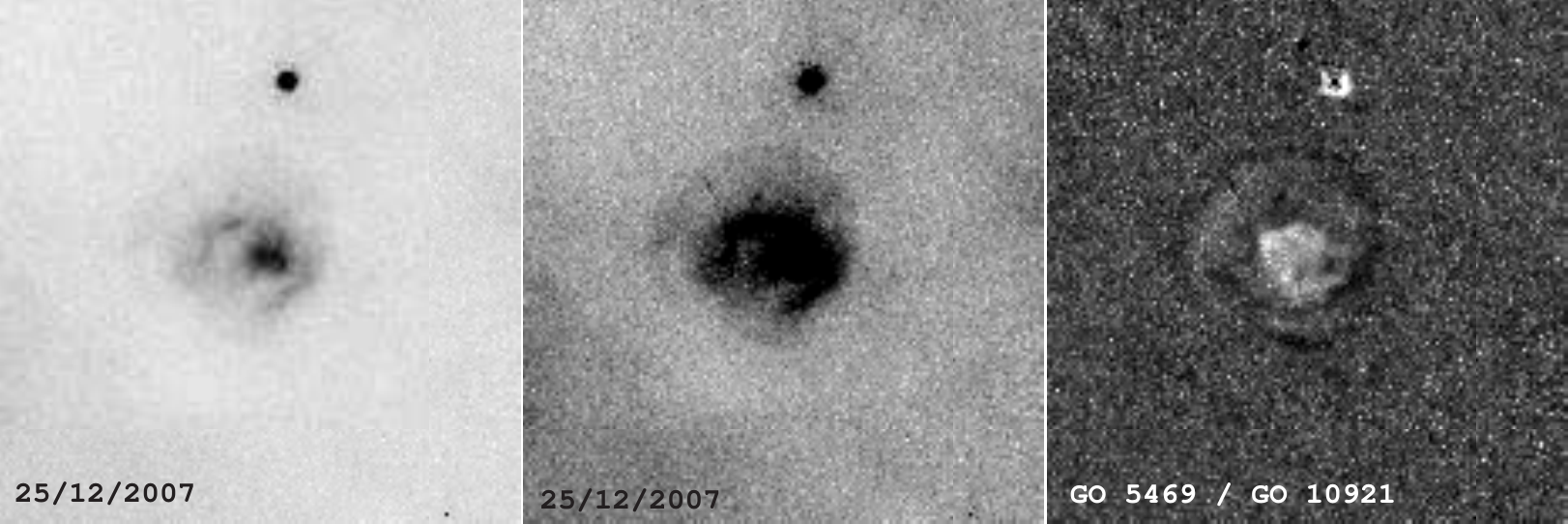}
\caption{This \(7.1\arcsec \times 7.1\arcsec\) view of HH 208 shows the structure and changes in the object.
The left two panels are brightness inverted views of the inner and outer regions of the object in the GO 10921 F673N ([S~II]) PC image (0.0455\arcsec /pixel). The right hand panel is a positive depiction of 
the ratio of earlier to recent images, showing the changes of structure in the inner region and general expansion in the outer. The nearby star is 137-222=JW~420\citep{jw85} and the vertical axis points towards 
PA=45\arcdeg.
\label{HH208ThreeView}}
\end{figure}

It is worth asking whether the optical features can help to illuminate what is happening.
Many of the molecular fingers (\htwo\ is the best way of tracing them) have low ionization optically visible shocks at their tips (HH 201, HH 210, and the HH 205+HH 206+HH 207 combination are the best examples), with HH 201 being the brightest and best studied \citep{gra03,doi04}. The resolved form of the shocks reflect their oblique angle to the viewer, but one of them, HH 208, appears closest to the origin, is seen almost along the line of sight, and fell within the high resolution (0.0455\arcsec / pixel)field of view in our GO 10921 observations, having previously been imaged in the lower resolution (0.0966\arcsec / pixel) WFPC2 CCD's in programs GO 5469 and GO 8121 \citep{ode97}. It is undetectable against the nebula in the \Halpha, [N~II], and [O~III] filters but stands in good contrast in the [S~II] and [O~I] filters, especially [S~II].  There is no discernible motion of this object, the centroid  (5:35:13.59 \(-5\):22:23.9) of this irregular object being constant to about 3 \kms.  What is most noteworthy is the change of structure that has occurred. Figure 8 shows the present appearance both in the core (middle panel) and the outer parts (left panel), while the right panel shows the ratio images between the recent (GO 10921) and oldest (GO 5469) observations (12.76 years), after magnifying to the same resolution.  The changes in structure are remarkable, well beyond what might be attributed to sampling with different size pixels, and argue that the form of this nearly head-on bow shock is highly variable.  The outer parts of it are expanding at about \(\Vt = 48~\kms\), yielding a dynamical timescale of 70 years. We have determined the radial velocity of the object from [S~II] spectra \citep{gar08} and found \(\Vsun = -165~\kms\), high, but much lower than in HH 201 and HH 210 \citep{hu96}.   These numbers indicate that the motion is within 1\arcdeg\ of our line of sight and it must be nearly in the same direction as its source.

\begin{figure}
\epsscale{1.0}
\plotone{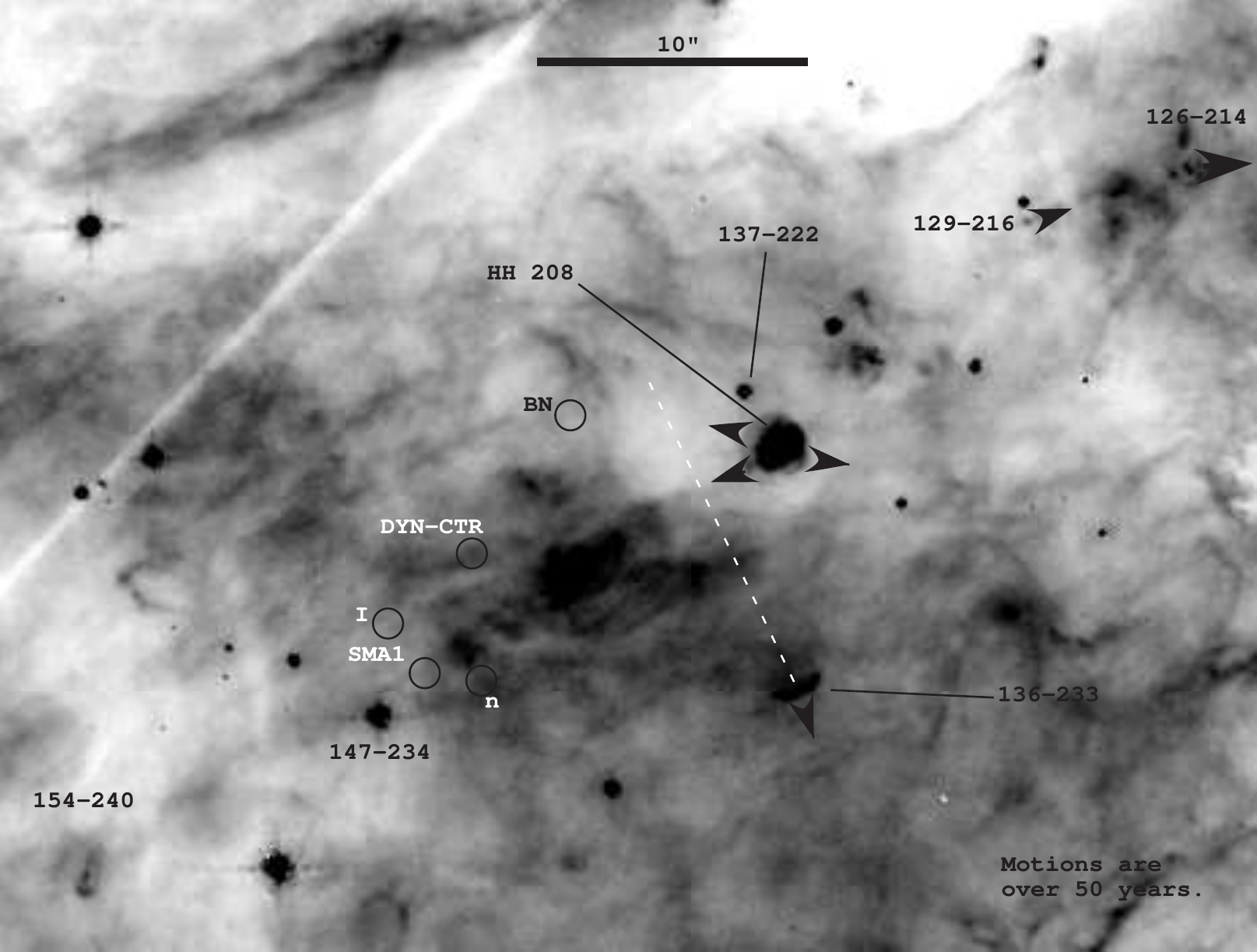}
\caption{This 46.3\arcsec x 35.1\arcsec\ WFPC2 F673N inverted brightness image at 0.0996\arcsec /pixel from program GO 10921
shows the region near HH 208 with north up. DYN-CTR indicates the position of the dynamical center of expansion of the radio sources \citep{gom05} and the nearby sources BN, I, and n. The submm source SMA1's position is taken from \citet{beu08}. The motions of objects 126-214 and 129-216 are from \citet{doi02} and the others are from this study. The dashed line indicates the axis of symmetry of the shock 136-233.The light diagonal band in the upper left occurs at the juncture of two of the WFPC2's detectors.
\label{BNKLregion}}
\end{figure}

A wider field of view around the putative source is shown in Figure 9.  There is a series of low ionization features moving away from the source region, two knots within them having measured tangential motions (timescale 900 years for motion from the DYN-CTR point).  However, this region is very complex, there being a linear array of fine features passing from the proplyd 154-240 \citep{schu99}, passing through the star image 147-234 and SMA1, south of DYN-CTR, and stopping at the bright knot of [S~II] emission, short of reaching HH 208, but pointing towards the shocks containing 129-216 and 126-214. 
There is a well defined shock (136-233) moving towards PA=198\arcdeg, and it has an \htwo\ feature on its leading edge \citep{schu99}.  Its symmetry axis does not point back to any of the candidate sources.  If it began at DYN-CTR, the timescale would be 700 years at \Vt =38 \kms, so that again one finds a dynamic timescale for the motion that is about the same as the outer HH objects and the central dynamical event. Repeated observations of this region with WFPC2 will not be possible since that instrument will be replaced during servicing mission SM4 (now scheduled for October, 2008), but repeat observations in \htwo\ \citep{schu99}  using the NICMOS instrument may be able to use tangential velocities to narrow down the source of the outflow.

\subsection{Outflows in the HH 502 Region}

\begin{figure}
\epsscale{1.0}
\plotone{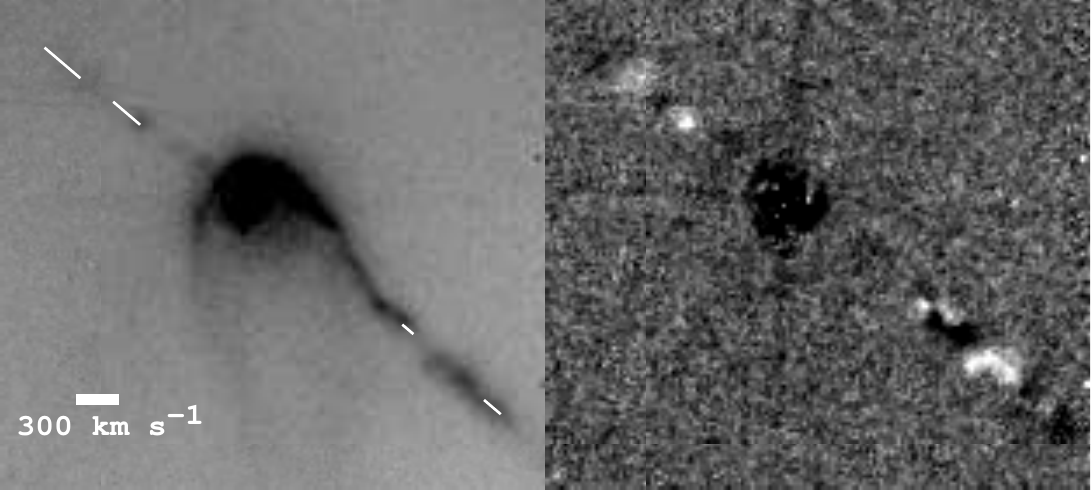}
\caption{
The central \(7.9\arcsec \times 7.1\arcsec\)  region around the proplyd driving HH 502 is shown with PA=-23.4\arcdeg\ as up. The left panel shows the object in inverted brightness and the right the ratio of the first-epoch/second-epoch images. The white lines indicate the measured motion away from the central star over the observation interval of 4.29 years. The upper left knots have a dynamic age of  16$\pm 2$ years and the lower right knots a dynamic age of 56$\pm 4$ years.
\label{HH502core}}
\end{figure}
\begin{figure}
\epsscale{1.0}
\plotone{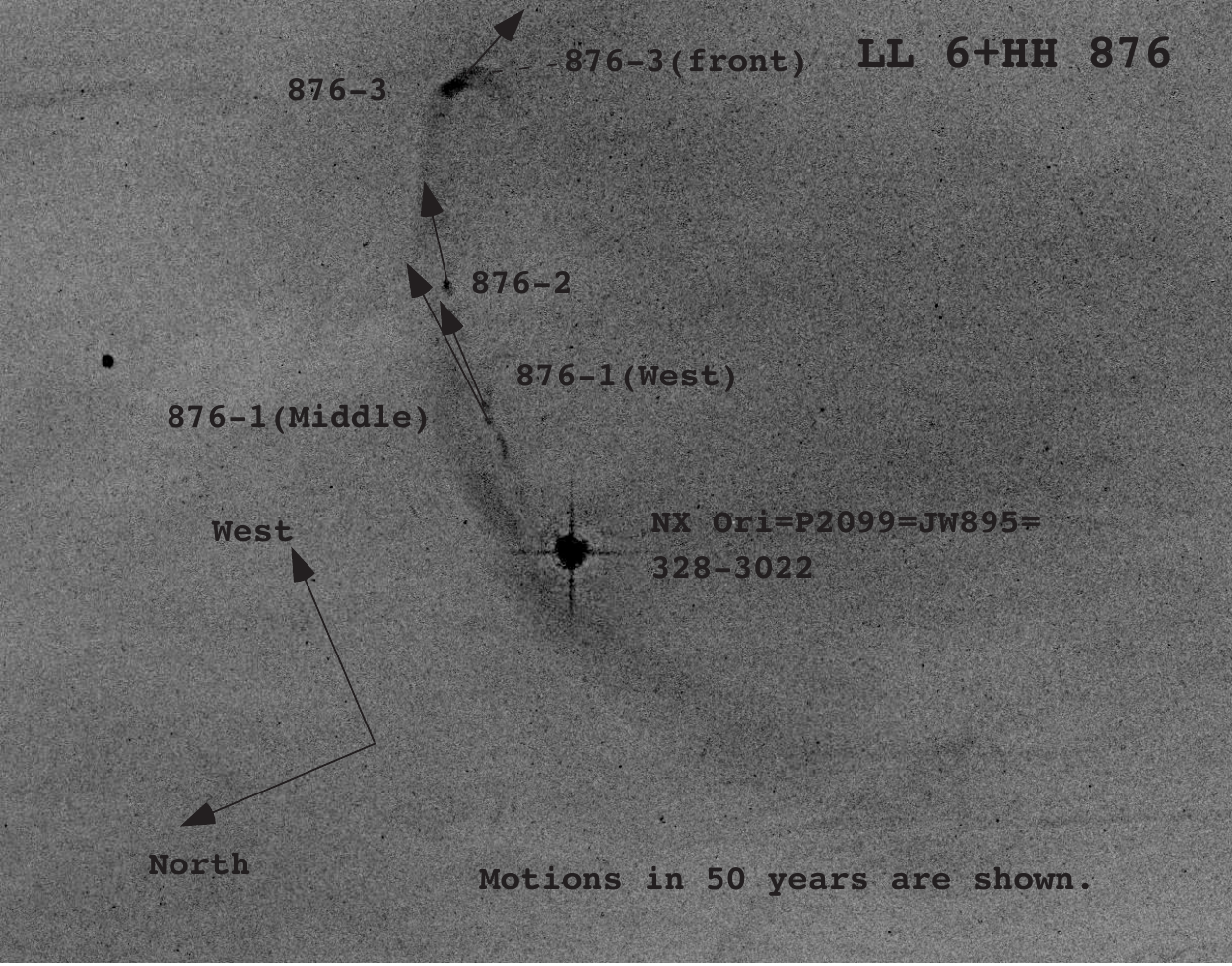}
\caption{
This inverted brightness, \(60.9\arcsec \times 47.6\arcsec\), image of LL6 and its driving jet (HH 876) in the ACS-WFC F658N filter shows the motions in the jet and the knots in the curved shock characteristic of the LL Ori class of objects. 
\label{LL6}}
\end{figure}

The area south of the Huygens region is known to be very rich in bow shocks and outflows \citep{br01,bal01,bal06,hen07}. This is probably because the luminosity of a shock is primarily dependent upon the density and velocity of the outflowing material, which must be basically independent of distance from \ori, whereas the confusing background brightness of the nebula has diminished because of the larger distance from the source of ultraviolet photons that ultimately determine the emission-line brightness of the nebula. The region centered on HH 502 has been thoroughly studied by \citet{bal06} using ACS-WFC F658N images taken 1.465 years apart. In the present investigation the images were made with a separation of 4.293 years, hence the tangential velocity resolution was superior, being 24.3 \kms per 0.05\arcsec\  per pixel. The larger field of view over which we determined tangential velocities is shown in Figure 6 and smaller enclosed fields around the proplyd 280-931 and LL 6 (HH 876) are shown in Figure 10 and 11.  We adopt the system of nomenclature of \citet{bal06}, which suffices as only a few new features were measured. In fact, the tangential velocities are typically high enough that our longer time-base observations produce results (Appendix C) that are very similar to those in \citet{bal06} with the exception of object HH 502-S7, where our value is about half of that of \citet{bal06}.

\paragraph{HH 502.} The most striking system in the measured field is HH 502, the bipolar flow discovered from groundbased observations \citep{br01}, which in form very much resembles HH 513 in the Huygens region both in having quite broad shocks, but also in showing a slight curvature away from \ori\ with increasing distance from the originating proplyd 280-931.  

One observes the structured jets that  drive both outflows as one approaches 280-931. Figure 10 shows both the first epoch appearance of this inner region and also shows in a ratio image the changes in structure and movement that has occurred in even the relatively brief interval of the observations we have used.  The dynamic timescale for these objects are 15.1 years (1), 17.2 years (2), 54.8 years (14), and 56.9 years (15), giving us the shortest time resolution of outflow events in the Orion Nebula.  The timescales for the outer features are much longer, being N3 (4490 years), N2 (1790 years), N1 (87 years), S3 (460 years), S4 (420 years), S5 (850 years),  S6 (1550 years), and S7 (3430 years). 

\paragraph{HH 876-LL 6.}
This object closely resembles the archetype for these objects LL Ori \citep{bal00,bal06,hen07} in having a broad bow shock that is indistinguishably stationary facing the flow of gas in its region, and a nearly perpendicular bipolar jet that produces its own set of rapidly moving small bow shocks. In this case the bow  shocks have the short timescales of 275 years (HH 876-3), 140 years (HH 876-2), and 56 years (HH 876-1).   

\subsection{HH 540}

The ``Beehive'' proplyd (181-826) HH 540 and its surrounding field has been imaged with the ACS-WFC in the F658N filter and thoroughly discussed in \citet{bal05}.  There is an inner silhouette disk with orthogonal microjets associated with shocks along slightly curved trajectories to both the north and south. However, the most remarkable thing about the object is the series of rings along the bright shell of the proplyd, these being symmetric about the bipolar flow axis, suggesting the nickname Beehive as it resembles a beehive as rendered in old cartoons.  The cause of these rings is unknown, but \citet{bal05} have probably identified the most likely cause. The jet that penetrates the ionized portion of the proplyd must be irregular in flow, thus providing a variable stimulus for perturbations that would then pass along the local ionization front. \citet{bal05} point out that a single similar ring is seen in the proplyd 182-413. \citet{bal05} calculate propagation velocities for the perturbations from the dynamic ages of the distant shocks driven by 181-826's outflow, finding velocities of a few kilometers per second and linking this to sound speed in the gas just inside the ionization front of the proplyd. It seems equally likely that the rings one now sees are produced by the much closer (and younger) knotty structure in the outflow, which would yield higher velocities of the rings and arguing that they are perturbations propagating in the ionized portion of the proplyd.

\begin{figure}
\epsscale{0.5}
\plotone{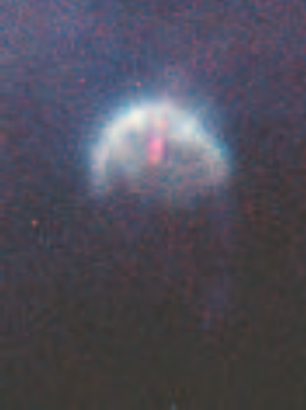}
\caption{
This 8.8\arcsec x11.8\arcsec\ image of HH 540 made with the WFPC2 is color coded as described in the text. The vertical axis points toward PA=355.3\arcdeg.  The inner silhouette disk is discernible around the central star as are two microjets, the lower (southern and opposite to the ionizing star \ori ) being brighter. The ripples in the ionized shell are probably caused by irregularities in the rate of mass loss within the microjets.
\label{HH540}}
\end{figure}

Our WFPC2 images cannot improve on the clarity of the view of HH 540, however, our observations included images in a wide range of emission line filters (F502N, F656, F658N, and F673N) in addition to the medium-width filter F547M . This allows us to discuss the ionization structure better, especially since the ACS-WFC F658N used in the earlier study passes both \Halpha\ and [N~II] emission equally well, thus leaving an ambiguity about what one is actually observing.  We have created a color image (Figure 12) showing F656N (\Halpha) as blue, F658N [N~II] as green, and F673N ([S~II]) as red. There is the unusual situation that the microjet is more visible on the side away from \ori. This is especially true in [S~II] emission, which indicates that this side is shielded by the inner disk from the FUV radiation that can pass through the proplyd's ionization front, whereas \citet{bal05} explain the greater brightness in F658N on the south side as being due to there being less obscuration by the disk on that side due to the orientation.  It is more likely that the EUV illumination process is more important as the south side is the blueshifted.  One also sees in the color image that there seems to be a progression of ionization states in the rings.

We have used the superior emission-line isolation to calculate the surface brightness of the proplyd in \Halpha, using the technique and calibration constants presented in \citet{ode99}.  The full chord width is 3.5\arcsec, the distance from the central star to the bright cusp is 1.0\arcsec, and the distance from the bright cusp to the end of the tail opposite \ori\  is 6.3\arcsec. \citet{bal05} indicate that the proplyd is 303\arcsec\ from its photoionizing star \ori. Using the definition of \citet{ode99} for the average surface brightness (their \S\ 3, paragraph 3), we find this to be \(8.3 \times 10^{8}\) \Halpha\ photons cm$^{-2}$ s$^{-1}$ ster$^{-1}$. A direct comparison with the calibration in \citet{bal05}, where the \Halpha\ line could not be isolated, indicates that their surface brightness is about 1.7 times too small.

We can compare the results for this large distant proplyd with similar objects found closer to \ori\ by reference to a study of the Huygens region proplyds \citep{ode98}.  The chord size fits right into the middle of the pattern of size versus distance. The surface brightness is 0.14 times (dex=0.85) that predicted by the upper limit in the surface brightness-distance relation. This relation would only be satisfied if the proplyds all lie in the same plane as \ori\ and all of the ionizing Lyman continuum photons were absorbed at the ionization front of the proplyd. The former condition is not satisfied, causing the Huygens region extinction corrected proplyds to run about 0.35 dex lower than the maximum brightness prediction. Reconciling the observed surface brightness and that expected from the properties of the Huygens region proplyds requires an extinction of 0.50 dex at \Halpha. We do not know the extinction in this region, but a value of 0.50 dex from the foreground veil is similar to that in the outer regions where it has been determined \citep{ode00}.  The extinction would be even less if the proplyd lies more than an average amount outside the plane containing  \ori. These comparisons indicate that the Beehive proplyd quantitatively behaves like the inner region proplyds in spite of its much greater distance from \ori\ and its bizarre appearance.

\subsection{New, Expanded, and Newly Clarified Herbig Haro Objects}
The availability of this new set of data on tangential motions and the high resolution ACS-WFC images has allowed the discovery of several new outflows, understanding new detail in a few HH objects, and has caused rethinking about what composes two already designated objects.

\begin{figure}
\epsscale{1.0}
\plotone{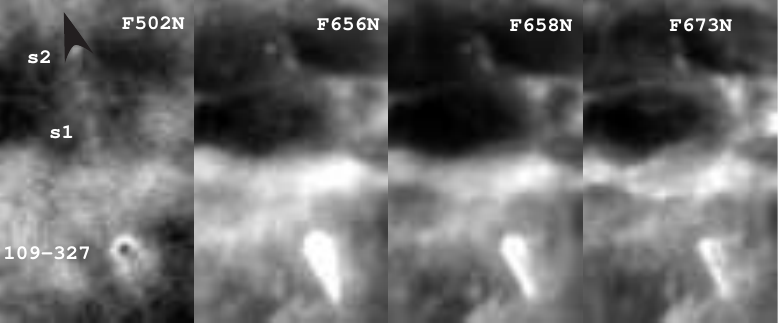}
\caption{
This series of 5.6\arcsec x 9.3\arcsec\ WFPC2 images of HH 510 shows the wide variation in appearance in different emission line filters. PA=46\arcdeg\ is up. The arrowhead indicates motion over 20 years.
\label{HH510}}
\end{figure}

\paragraph{HH 510.} This proplyd (109-327) was discovered \citep{bal00} to have a microjet on a wide bandpass (F606W) WFPC2 image and designated as HH 510. The short microjet extended in the direction of two sharp features (s1 and s2 in Bally et al.'s Figure 15 and Table 3). Our new  observations were able to measure motions in the brighter s2 feature and to determine that it is at the tip of the high ionization jet (the s2 feature is visible in \Halpha, [N~II], and [O~III]). This knotty jet can be traced all of the way (5.9\arcsec) to s2 (Figure 13).  Feature s2 is moving at 101 \kms\ towards PA=64\arcdeg. The timescale for this motion is only 120 years and presumably much shorter for the knots closer to the proplyd.

\begin{figure}
\epsscale{1.0}
\plotone{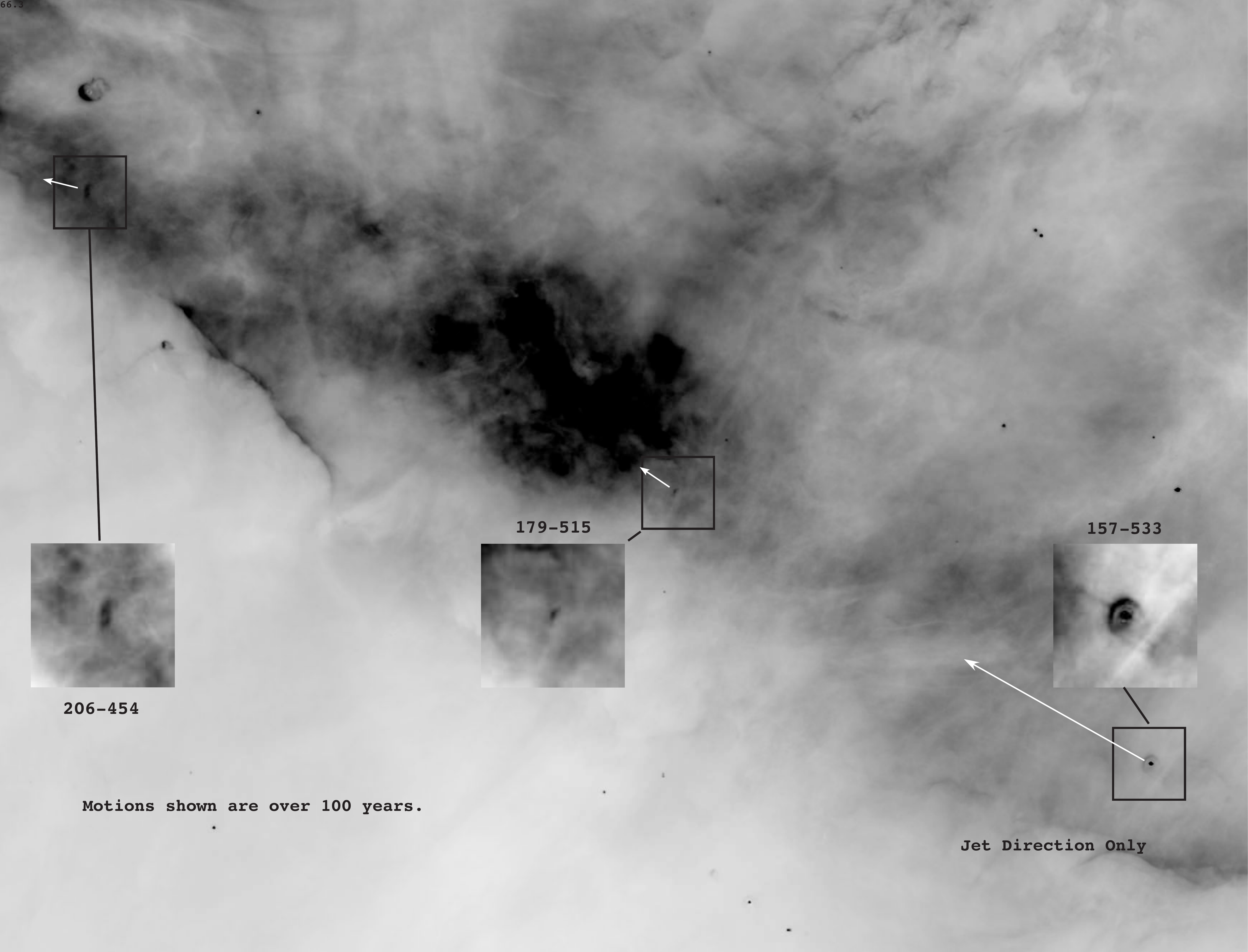}
\caption{
This inverse brightness depiction in the ACS-WFC F658N (\Halpha + [N~II], \citep{hen07}) filter shows an \(86.9\arcsec 
\times 66.3\arcsec\) field near the head of HH 528(large dark feature in the middle) with north up. The insets show enlarged \(10\arcsec \times 10\arcsec\) regions. The proplyd 157-533 and its microjet has previously been designated as HH 512 \citep{bal00}. The alignment with this microjet, their forms, and their motions all indicate that the shocks 179-515 and 206-454 are part of this HH object. 
\label{HH512}}
\end{figure}

\paragraph{HH 512.} The proplyd 157-533 (Figure 14) was previously known \citep{bal00} to have a microjet projecting towards PA=60\arcdeg, which lead to its designation as HH 512. Examination of the GO 10246 ACS-WFC F658N image shows there to be a very small (0.9\arcsec\ length) inner silhouette disk indistinguishably perpendicular to the microjet and that the microjet extends to 3.5\arcsec\ from the central star. A shock (179-515) within the ``cap'' of HH 528 was already known \citep{doi02} to possess an anomalous motion (we now measure this to be 104 \kms\ towards PA=56\arcdeg), but a connection with HH 512 was not made. We now note that this bow shock lies along the projection of 157-533's microjet and is moving away from that proplyd.  Our reprocessing of the images made near HH 203 and HH 204 show an additional bow shock (206-454) lying almost on the line between 157-533 and 179-515 and is again moving away from the source (103 \kms\ towards PA=76\arcdeg) and is almost certainly a component of this HH object. The timescales are comparable to those of many other Huygens region HH objects, being 770 years (206-454) and 1700 years (179-515). The remarkable thing about these bow shocks is that their small size is evidence that the driving jet, which is essentially unresolved near the source, has not expanded to more than 1.4\arcsec\ at a distance of 84.3\arcsec.

\paragraph{HH 513.} This well studied \citep{bal00,doi02} (Figure 3, Figure 4, and Figure 5) object is the only  bipolar object in the Huygens region where one can measure tangential motions on both sides of the outflow.  Our observations indicate that the east side components are moving at about 34 \kms\ and those on the west side at 75 \kms, indicating timescales of 100, 200, 240, 310, 430, 1500, and 2700  years for the seven component shocks.  We have measured [O~III] spectra  for the west flow from the dataset created by \citet{doi04} and determined that \(\Vsun = -30~\kms\), indicating that the west side is the closer side and that it is at an angle of about 37\arcdeg\ with respect to the plane of the sky.

\begin{figure}
\epsscale{1.0}
\plotone{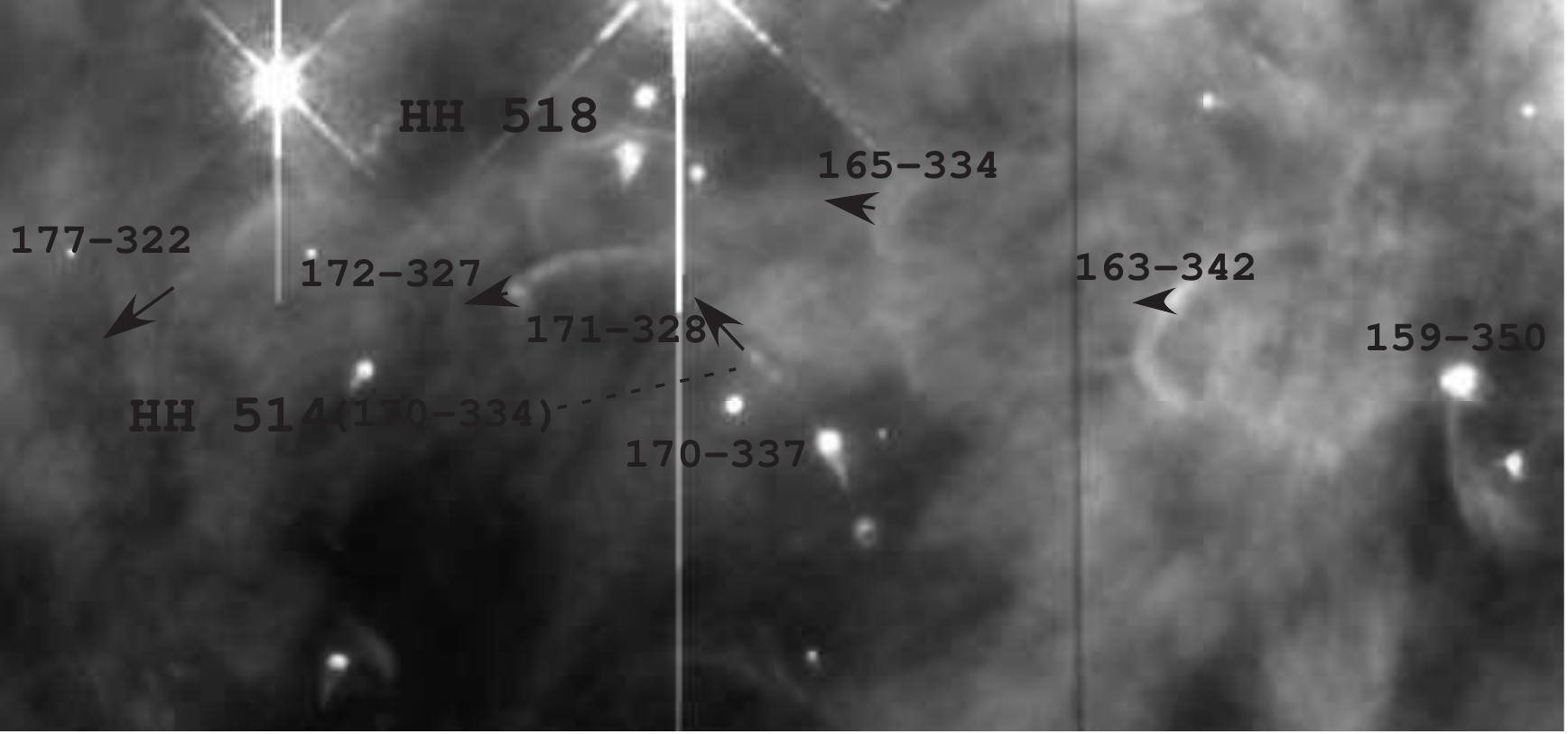}
\caption{
This 49.7\arcsec x23.2\arcsec\ image of the components of HH 514 and HH 518 uses the same WFPC2 images as Figure 5 but with a greyscale rendering and has PA=314.4\arcdeg\ as up.  The bright star just off the field of view is \ori. The length of the arrows indicate tangential motions in 100 years.  
\label{HH518}}
\end{figure}

\paragraph{HH 514 and HH 518.} These join HH 873 \citep{gar07} as the only redshifted Herbig Haro outflows in the Huygens Region.
The features included in these two HH objects have been different in the several articles that have considered them \citep{bal00,doi02,od03}. Fortunately, our new data allow clarification of this rather complex region, which is shown in Figure 15. 
The simplest system is HH 514, which begins as a redshifted microjet arising from the proplyd 170-337 (there is a faint blueshifted counter jet in the opposite direction) and continues with a group of rapidly moving knots (170-334) about 3\arcsec\ north of the proplyd \citep{bal00}.  \citet{bal00} then say that there are a series of associated shock features at 5.6\arcsec, 8.9\arcsec, 12.9\arcsec, and 14.6\arcsec.
Only the feature at 5.6\arcsec\ (which is visible on our images but its motion was not measured) is probably part of HH 514 unless there is a very abrupt change in direction of the flow.  \citet{bal00} define HH 518 as a single shock (designated here as 163-342) associated by symmetry and proximity with the proplyd  159-350 and say that there were other fainter associated shocks further from the source, without making clear identifications. \citet{doi02} argued that the shock at 172-326 (172-327 here) was a new HH object, calling it HH 605. However, with only little curvature, it should be identified as part of the HH 518 flow \citep{od03}. We believe that HH 518 is composed of the four shocks 177-322, 172-327, 165-334, and 163-342 for which we have determined tangential velocities.  The higher ionization radial velocity mapping of this region \citep{doi04} provide radial velocities of \Vsun = 98 \kms, 104 \kms, 74 \kms, and 102 \kms\ respectively.
This means that HH 518 is also anomalous within the Huygens region in being a redshifted flow.  Moreover, examination of the \Halpha\  spectrum 159-350 \citep{doi04}  shows a redshifted jet. The values of $\theta$\ derived from our tangential velocities and these radial velocities are \(-51\arcdeg\), \(-69\arcdeg\), \(-55\arcdeg\), and \(-77\arcdeg\), that is, the flow is highly inclined and pointing away from the observer. 
A very similar result applies for HH 514, where our tangential motions and the radial velocity (\Vsun =152 \kms) from \citet{bal00} give \(\theta=-69\arcdeg\).  The unusual and similar values for these flows argues that there is something different and possibly related in the outflows from 159-350 and 170-337. 

\paragraph{HH 626.} This object is an incomplete double ellipse (possibly two opposing systems of broad bow shocks)  centered on the proplyd 143-425 \citep{od03}. Our tangential velocities indicate timescales of 450 and 500 years for the northern components and 360 and 750 years for the southern components (excluding a single northern component of 1900 years, which is probably due to an error in the \Vt = 8 \kms\ determination).  

\begin{figure}
\epsscale{1.0}
\plotone{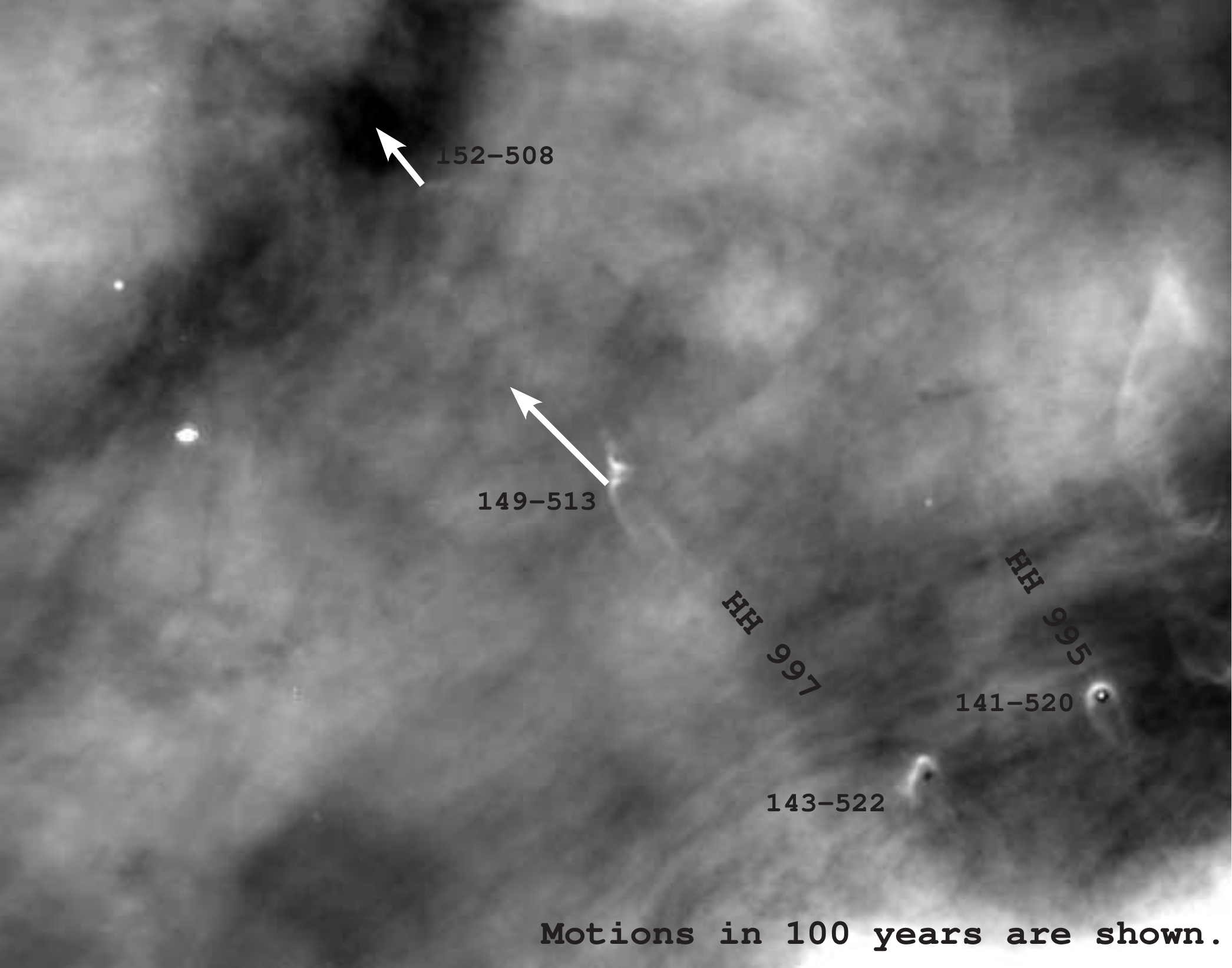}
\caption{
This \(30.1\arcsec \times 23.7\arcsec\) ACS-WFC F658N image (north up) shows the newly designated HH 995 and HH 997. HH 997 originates in the dark-disk proplyd 143-522, where one sees a faint microjet extending to the northeast, perpendicular to the dark disk and towards the two associated shocks. The middle shock (149-513) is centered on a narrow linear feature that is likely to be an extension of the microjet. The microjet forming HH 995 arises from 141-520.
\label{HH997}}
\end{figure}

\paragraph{HH 995 and HH 997.} Our ratio images indicate a pair of shocks (152-508 and 149-513)  oriented towards PA=45\arcdeg\ and moving towards PA=51\arcdeg. The westward of the pair has a three arcsecond linear feature lying along its symmetry axis (Figure 16).  A projection backwards through these two shocks passes close to the newly discovered  bright proplyd 143-522 with a silhouette dark disk, which shows a microjet extending (PA=45\arcdeg) towards the shocks and is almost certainly their driving source. There is another newly discovered similar object (141-520) with a microjet extending towards PA=40\arcdeg\ which has been designated as HH 995. The timescales for movement from 143-522  are 1100 years (152-508) and 300 years (149-513).

\begin{figure}
\epsscale{0.8}
\plotone{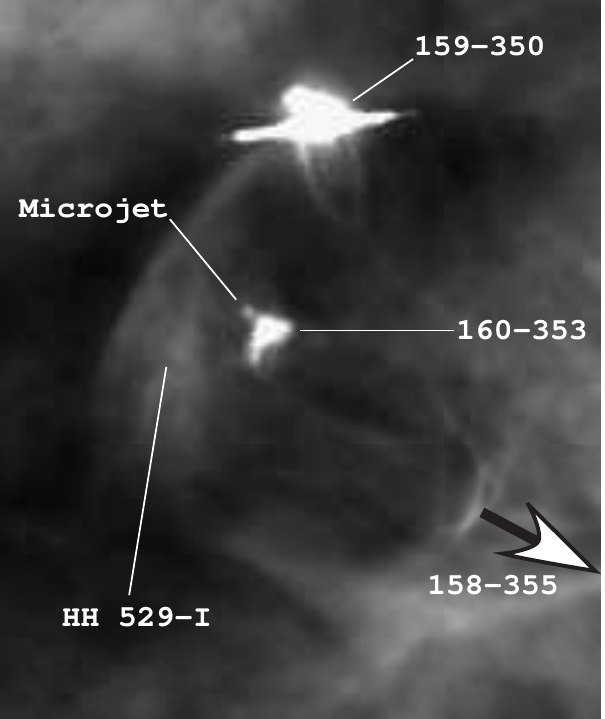}
\caption{
This \(8.65\arcsec \times 10.35\arcsec\) ACS-WFC F658N image (north up) shows the newly discovered microjet extending to the northeast from the proplyd 160-353 and the shock 158-355 which is in the opposite direction of the microjet. This ensemble is newly designated as HH 998. HH 529-I is the first of a series of large, high ionization, eastward moving shocks within the system designated as HH 529. The bright feature extending to the northeast from 159-350 is an overexposed image of a companion star, not another microjet. The arrow designates the motion in fifty years.
\label{HH998}}
\end{figure}

\paragraph{HH 998.} The bright rim proplyd 160-353 (Figure 17) reveals a new feature on the GO 10921 and the ACS-WFC
images in that it has a 0.4\arcsec\ microjet extending towards PA=47\arcdeg and nearly opposite to this microjet there is the rapidly moving (80 \kms) shock 158-355.  It is not unusual in Orion to find that the microjet is on the side facing the nearby photoionizing star (\ori\ in this case) and there is no reason to not expect a similar, but less visible, outflow in the opposite direction. It is this currently-invisible counter-jet that must be driving 158-355 and the timescale for the separation is only 100 years.

\subsection{Possible new HH Objects}

There are many features for which we have determined tangential velocities but these features have not been assigned to existing or new  (HH 995, HH 997, and HH 998) HH objects. In some cases there are indications of the need to assign them membership in a new HH object, but the level of evidence does not reach a sufficient level.
We mention several of these in this section.

\paragraph{Outflow from 117-352?}  Two nearly aligned  shocks (110-352 and 116-352) are moving away from the nearby proplyd 117-352 (Figure 5), however, this area overlaps with the large HH 269 shocks and they may be part of that HH object.

\paragraph{Outflow from 150-355?} There are two shocks (149.3-351.7 and 150.5-351.9) just north and moving away from the star 150-355 (Figure 7). This motion is quite different from other flows occurring in this area.

\paragraph{Outflow associated with the Dark Arc?} There are four low tangential velocity features (130-345, 131-342, 133-342, and 140-344) which appear (Figure 7) to be moving away from the geometric center of this unexplained object \citep{ode00}.

\paragraph{Motions near 143-353 and 145-355+145-356.} There are three southeast moving features (151-359, 154-401, and 158-405) aligned with both the infrared stars 145-355 and 145-356 and the more distant radio source 143-353). A northwest moving object (144-354) lies between the two possible driving sources. If all four objects  are part of the same bipolar flow, then the source would be either 145-355 or 145-356.  In the infrared polarization study of \citet{has07} one sees a polarization signature around the latter pair of stars. Although they favored 145-356 as the source, the low angular resolution cannot make a definitive distinction between the two. The direction of the southeast moving features points towards the high velocity jet that drives HH 203 and it is possible that these are all one system.

\paragraph{A Very High Velocity Feature.} 
One of the highest tangential velocity features found in this study is 137-317 (\(\Vt =155~\kms\), PA=113\arcdeg). It is very low contrast, only about 10\%\ brighter than the surrounding background and is only clearly discernible in F658N, barely detected in F631N and F673N, and not visible in F502N and F656N. 
It is a very low ionization shock of about 1.5\arcsec\ width, symmetric to the direction of motion. There are no candidate sources nearby for this unusual feature.
 
\subsection{General Properties of Stellar Outflows}

The particular advantage of studying stellar outflows from objects in the ONC is that one is dealing with a group of stars all about the same age and sharing similar histories.  This means that we can draw general conclusions that probably are applicable to other HH, which are usually found as isolated, single, objects.

A striking pattern of the HH objects in the ONC is the intermittent nature of the outflows, a not uncommon characteristic that was first documented on smaller scales by \citet{rei89} and on larger scales by \citet{rei97}. Our time resolution is probably best for the close shocks and jets because distant bow shocks can actually be the result of multiple close-together in time outflows. In single objects like HH 502 we find well defined variations in the bipolar outflow with time gaps of a few tens of years.  These staccato intervals of outflow are probably spaced out at larger intervals, producing the larger scale features that one sees at greater distances from the sources. This means that bipolar outflow is not at all continuous, rather, there is evidence that it occurs in concentrated periods at intervals of a few hundred years and during the period of outflow there is a irregular mass loss with a timescale of about ten years. Even shorter timescale variations may be present, but our observations would not have been able to detect typical outflows of shorter than a few years.

With the exception of HH 625, the largest scale HH objects in the Huygens region appear to originate in a small area, immediately northeast of the brightest Orion-S radio and infrared sources and they do not arise from the SiO and CO molecular outflows.  \cite{zap06} point out that the sources driving the known molecular outflows are not particularly luminous and it is not surprising that even less luminous and less obscured sources would be able to produce the highly visible objects like HH 202, HH 203, HH 204, HH 269, and HH 529.  The ubiquity of blueshifts in these HH objects  argues that these originate from different young stars, because linking them to a common source would require large changes in orientation in the plane of the sky and along the line-of-sight with timescales of a few hundred years.  

These observations do not give us definitive information about long timescale variation in the direction of the axis of outflows since the visible outflow material is moving through the ambient gas of the nebula (which will have a local low velocity flow as it moves away from the main ionization front and from the brightest part of the nebula) and, in some cases, will be affected by the stellar wind from the hot massive stars.





\acknowledgments

We are grateful to Yilen G\'omez Maqueo Chew, who helped us obtain the new spectroscopic observations, to Maria-Teresa Garc\'{\i}a-D\'{\i}az, who allowed use of her earlier spectra, to Massimo Robberto and Vera Platais, who provided the Orion ACS images in a user-friendly form, and to Bo Reipurth for maintaining a uniform catalog of HH objects and assigning new numbers to the ones discovered in this study. CRO's work was supported in part by the Space Telescope Science Institute grants GO 10921 and GO 10967. WJH acknowledges financial support from DGAPA-UNAM, Mexico
(PAPIIT IN110108).



{\it Facilities:} \facility{HST (WFPC2)}, \facility{SPM}.



\newcommand\tn{\tablenotemark}
\newlength\savetcs\setlength\savetcs{\tabcolsep}
\setlength\tabcolsep{3pt}
\begin{deluxetable*}{ccccl c@{\hspace{4pt}} ccccl}
\tabletypesize{\tiny}
\tablecaption{Tangential Velocities in the Huygens Region
  \label{tab:huygens}}
\tablewidth{0pt}
\tablehead{
\colhead{Designation} & \colhead{HH} & \colhead{\shortstack{\Vt\\ (\kms)}} & \colhead{\shortstack{PA\\ (\arcdeg)}} & \colhead{Filters} 
& \colhead{} &
\colhead{Designation} & \colhead{HH} & \colhead{\shortstack{\Vt\\ (\kms)}} & \colhead{\shortstack{PA\\ (\arcdeg)}} & \colhead{Filters} 
}
\startdata
100-246&202& 49& 302&F502N,F656N &                     &170-359&529-III& 41 &119 &F631N,F656N,F658N,F673N\\
104-245&202& 38& 298& F502N,F656N &                    &109-321&625& 25& 302 &F631N,F656N,F658N,F673N\\                 
112-246&202& 31& 310& F502N,F656N &                    &110-323&625& 39& 307 &F631N,F673N \\                            
115-255&202& 57& 308& F502N      &                     &111-324&625& 36& 289 &F631N\\                                   
116-250&202& 58& 336&F502N,F656N&                      &112-324&625& 37& 285 &F656N,F658N,F673N\\                       
116-252&202& 36& 343&F502N            &                &140-419&626&  6& 001 &F656N\\                                   
116-255&202& 80& 338&F656N,F658N,F673N &               &141-419&626& 29& 001 &F502N,F656N \\                            
116-256&202& 41& 305&F502N           &                 &141-421&626& 19& 343 &F502N,F656N\\                             
116-254&202& 52& 334&F656N,F658N    &                  &142-428&626& 10& 163 &F656N\\                                   
117-249&202& 36& 322&F658N,F673N   &                   &143-429&626& 26& 208 &F502N\\                                   
117-255&202& 33& 316&F631N         &                   &172-320\tn{c}&726& 98& 094 &F502N \\                            
116-344&269& 52& 265&F502N        &                    &149-513&997&70 &045& F656N,F658N\\                            
116-346&269& 38& 268&F502N,F656N &                     &152-508&997&37 &057&F656N,F658N\\                             
118-345&269& 61& 288&F502N,F656N&                      &158-355&998& 80& 242&F502N\\                                  
125-346&269& 38& 291&F656N     &                       &110-352\tn{d}&NEW(?)& 33& 302&  F658N,F673N  \\                   
137-351\tn{a}&269&16 &283 &F502N,F656N  &              &116-352\tn{d}&NEW(?)& 42& 269& F631N,F658N,F673N\\                
138-350\tn{a}&269& 33& 272&F656N,F673N  &              &149.3-351.7\tn{e}&NEW(?)&26&334& F656N\\                        
139-351\tn{a}&269&11 &302&F631N,F656N, F658N,F673N&    &150.5-351.9\tn{e}&NEW(?)&27&354& F502N,F656N\\                    
113-325\tn{b}&510&101&  64& F656,F502N,F658N&          &130-345&Dark Arc&20& 358& F502N,F656N \\                    
179-515&512&104& 056&F656N,F658N   &                   &131-342&Dark Arc&13& 316& F502N,F656N   \\                    
206-454&512&103& 076&F656N,F658N   &                   &133-342&Dark Arc&20& 295& F502N        \\                     
159-237&513& 59& 277&F502N,F658N&                      &140-344&Dark Arc&26& 026& F502N           \\                  
161-236&513& 60& 283&F658N      &                      &144-354\tn{f}&NEW(?)&50& 307& F673N \\                            
162-316&513& 98& 273&F502N,F658N&                      &151-359\tn{f}&NEW(?)&19&109& F673N  \\                            
168-235&513& 47& 076&F502N,F656N,F658N&                &154-401\tn{f}&NEW(?)&26& 154& F673N \\                            
171-234&513& 49& 053&F502N,F658N&                      &158-405\tn{f}&NEW(?)&35& 116& F631N,F656N,F658N,F673N\\           
173-233&513& 18& 035&F502N,F658N&                      &115-359& & 32& 278& F656N,F658N\\                             
177-229&513& 15& 025&F502N,F658N&                      &129-237& &  85& 351& F502N     \\                            
170-334&514& 48&357&F502,F656N,F658N  &                &131-246& &  9 &226& F658       \\                           
137-508&516& 33& 142&F631N,F673N       &               &132-343& & 17 &46& F502N       \\                           
163-342&518& 17& 046&F656N      &                      &133-344& &46 &1& F656N         \\                            
165-334&518& 34& 035&F502N,F656N &                     &134-332& & 23& 268& F631N,F658N,F673N\\                       
171-328&518& 33& 060&F502N   &                         &134-344& &  34& 1& F502N    \\                                
172-327&518& 27& 056&F502N,F656N,F658N&                &135-243& &29& 318& F502N    \\                                
177-322&518& 58& 080 &F502N            &               &135-347& &  22& 5& F502N,F658N\\                              
182-337&523& 74& 096 &F502N,F631N,F656N,F658N&         &136-233\tn{g}& &  38& 198& F673N\\                               
165-456&528& 20& 190 &F631N,F673N&                     &137-317& &155& 113& F658N\\                                   
166-434&528& 19& 196 &F631N,F673N&                     &115-359& & 32& 278& F656N,F658N \\                            
166-440&528& 26& 164 &F631N,F673N&                     &143-326& & 36& 289& F502N  \\                                 
166-442&528& 28& 203& F656N,F658N&                     &153-305& & 77& 334& F502N      \\                             
169-446&528& 20& 171& F656N,F658N&                     &155-253& & 13& 13 &F502N,F658N\\                              
170-459&528& 19& 163& F673N&                           &155-257& &  39& 334& F502N,F658N\\                            
171-502&528& 20& 161& F631N,F656N,F658N,F673N&         &157-326\tn{h}& &   41&305& F502N \\                              
177-454&528& 37& 199& F673N&                           &161-325& &   11&247& F502N,F658N\\                            
181-510&528& 26& 154& F631N,F673N&                     &161-410& &   26&208& F658N\\                                  
184-513&528& 18& 163& F631N,F673N&                     &165-317& &   25&316& F502N             \\                     
149.1-352.9&529&85&104& F502N,F656N,F658N,F673N&       &165-406& &   48& 96& F656N,F658N\\                            
149.6-353.4&529&95&105& F673N&                         &166-308& &  98 & 42& F502N\\                                  
150.1-353.3&529&91&105& F631N,F656N,F658N,F673&        &167-300& &   23& 38& F502N\\                                  
151.1-353.9&529&63&90& F502N,F631N,F656N &             &169-325& &  30 &125& F502N\\                                  
152-354&529& 18& 64& F673N&                            &171-218& &   17&  5& F502N,F658\\                             
155-354&529& 18& 91& F502N&                            &178-246& &   19&359& F658N\\                                  
156-356&529& 45& 109& F502N,F656N&                     &178-246& &   15&  6& F658N\\                                  
161-353&529-I& 30& 80& F656N&                          &178-306& &   20&200& F502N\\                                  
161-354&529-I& 78& 100& F502N,F631N,F658N,F673N&       &180-245& &  18 & 23& F658N\\                                  
167-358&529-II& 92& 91& F502N &    
\enddata
\tablenotetext{a}{Irregular Jet?\quad
\textsuperscript{b}{Feature S2 of \citet{bal00}}\quad
\textsuperscript{c}{Outflow from LV2's (167-317) jet?}\quad
\textsuperscript{d}{Driven by proplyd 117-352?}\quad
\textsuperscript{e}{Driven by 150-355?}\\ \hspace*{5pt}
\textsuperscript{f}{Driven by 144-355?}\quad
\textsuperscript{g}{Driven by BN(?), c.f. text.}\quad
\textsuperscript{h}{Symmetric with LV6 (158-326)}
}
\tablecomments{
Objects are grouped by assigned Herbig Haro number or potential new grouping. When there is no candidate Herbig Haro flow designation, they are listed individually. Within each category objects are ordered in increasing Right Ascension.}
\end{deluxetable*}

\setlength\tabcolsep{\savetcs}
\begin{deluxetable*}{lccl}
\tabletypesize{\scriptsize}
\tablecaption{Tangential Velocities near HH 502 \label{tab:hh502}}
\tablewidth{0pt}
\tablehead{
\colhead{Bally Designation} & \colhead{Tangential Velocity (\kms)} & \colhead{Position Angle (\arcdeg)} & \colhead{Comment}}
\startdata
HH 876-1(west) & 117 & 274&  Divided into two components.\\
HH 876-1(middle) & 67 & 270 & Divided into two components.\\
HH 876-2 & 64 & 260 & \\
HH 876-3 & 53 & 205 & \\
HH 560S-1 & 14 & 303 & \\
HH 560S-2 & 33 & 310 & \\
HH 560S-3 & 48 & 300 & \\
HH 502-1 & 248 & 27 & \\
HH 502-2 & 329 & 27 & \\
HH 502-N2(8) & 56 & 63 & \\
HH 502-N2(7) & 79 & 118 & \\
HH 502-N2(6) & 106 & 98 & \\
HH 502-N3(10-11-12) & 30 & 24 & \\
HH 502-N3(NE) & 48 & 22 & Additional feature measured.\\
HH 502-N1(5) & 76 & 43 & \\
HH 502-N1(3) & 224 & 35 & \\
HH 502-N1(4) & 182 & 35 & \\
HH 502-S3 & 107 & 198 & \\
HH 502-14 & 102 & 207 & \\
HH 502-15 & 153 & 207 & \\
HH 502-S4(16) & 158 & 202 & \\
HH 502-S5(17) & 122 & 199 & \\
HH 502-S6(19) & 100 & 197 & \\
HH 502-S7(21)& 54 & 196 & \\
HH 874(22) & 123 & 126 & \\
HH 874(23) & 154& 116 & \\
HH 874(24) & 168 & 157 & \\
HH 874(25) & 55 & 165 & \\
HH 874(26) & 58 & 164 & \\
\enddata
\tablecomments{The Bally Designations are those given in Table 3 and Table 4 of \citet{bal06}.}
\end{deluxetable*}

\begin{appendix}
\section{The Distance to the Orion Nebula}
The exact distance to the Orion Nebula Cluster eludes determination because its distance is too great 
to simply apply traditional parallax methods and it is ill-advised to apply indirect methods \citep{mue08,ode08}.
The pre-main Sequence stars are not suitable because of the many variables in determining their properties and the main sequence O and B stars are the closest exemplars, so that neither group can be used for an accurate spectroscopic parallax. \citet{jef07} has determined a distance using the statistical properties of the rotation of the cluster stars, but this result is very model dependent.

Trigonometric parallaxes are the fundamental method for determining distances, but even the Hipparchos satellite optical results are not definitive. \citet{ber99} noted one member star (HD 37061) and found a distance of 361$^{+168}_{-87}$ pc, while \citet{wil05} used multiple stars in this direction and by grouping according to if they are reddened by OMC-1 found 465$^{+75}_{-57}$ pc.

A recent optical interferometric study of \ori\ derived a distance as a by-product of the study of this star as a binary \citep{kra07}. They found two equally good solutions, Orbit 1 (434$\pm$12 pc) and Orbit 2 (387$\pm$11 pc). They favored the Orbit 1 result, but only because of better agreement with previous determinations.

Radio interferometric parallaxes are more promising.  \citet{gen81} applied what is in-effect the moving cluster parallax method to H$_{2}$O masers and derived a value of 480$\pm$80 pc. More recently, radio trigonometric parallaxes have been reported.  \citet{hir07} have studied a H$_{2}$O maser in BN-KL, determining its position in two dimensions from 15 observations and deriving a distance of 
437$\pm$19 pc. \citet{san07} measured a non-thermal source (GMR-A) in the inner ONC and found a distance of 389$^{+24}_{-21}$ pc. However, their source is highly variable in flux, is partially resolved, (it also shows variations in its structure), and both of the reference sources are resolved. In addition, the radial velocity of GMR-A is \(\Vsun =14\pm5~\kms\), which is quite different from the value (\(\Vsun =25.6~\kms\) \citealp{fur08}) for the stars in the cluster. The most recent study is that of \citet{men07} who measured only the Right Ascension
component of the parallax of four non-thermal sources (including GMR-A). The observations were made at only four epochs and one of the sources was not detected during one epoch, so that parallaxes were
determined for only three sources, finding 417$\pm$7 pc. If GMR-A is excluded, then the remaining two sources give 412$\pm$6 pc. 

We have derived a distance of 436$\pm$20 pc by assigning weight 3 to the Hirota et al.\@ 2007 study, weight 2 to the Menten et al.\@ (2007) and Kraus et al.\@ (2007) studies (Orbit 1), and weight 1 to the Genzel et al.\@ (1981) studies. We adopt the rounded value of 440 pc in this paper.

\section{Tangential Velocities in the Huygens Region}

Insert Table~\ref{tab:huygens} here.

\section{Tangential Velocities near HH 502}

Insert Table~\ref{tab:hh502} here.
\end{appendix}







\clearpage
\begin{thebibliography}{}
\bibitem[Bally et al.(2001)]{bal01} Bally, J., Johnstone, D., Jocas, G., Reipurth, B., \& Mall\'en-Ornelas, G. 2001, \aj, 122, 1508
\bibitem[Bally et al.(2005)]{bal05} Bally, J., Licht, D., Smith, N., \& Walawender, J. 2005, \aj, 129, 355
\bibitem[Bally et al.(2006)]{bal06} Bally, J., Licht, D., Smith, N., \& Walawender, J. 2006, \aj, 131, 473
\bibitem[Bally et al.(2000)]{bal00} Bally, J., O'Dell, C. R., \& McCaughrean, M. J. 2000, \aj, 119, 2919
\bibitem[Bally \& Reipurth(2001)]{br01} Bally, J., \& Reipurth, B. 2001, \apj, 546, 299
\bibitem[Bertout et al.(1999)]{ber99} Bertout, D., Robichon, N., \& Arenou, F. 1999, \aap, 352, 574
\bibitem[Beuther \& Nissen(2008)]{beu08} Beuther, H., \& Nissen, H. D. 2008, \apjl, in press
\bibitem[Blagrave et al.(2006)]{bla06} Blagrave, K. P. M., Martin, P. G., \& Baldwin, J. A. 2006, \apj, 644, 1006
\bibitem[Cant\'o \& Raga(1996)]{can96} Cant\'o, J., \& Raga, A. C. 1996, \mnras, 280, 559
\bibitem[Doi et al.(2002)]{doi02} Doi, T., O'Dell, C. R., \& Hartigan, P. 2002, \aj, 124, 445 (DOH)
\bibitem[Doi et al.(2004)]{doi04} Doi, T., O'Dell, C. R., \& Hartigan, P. 2004, \aj, 127, 3456 (DOH04)
\bibitem[F\"ur\'esz et al.(2008)]{fur08} F\"ur\'esz, G., Hartmann, L. W., Megeather, S. T., Sventgyorgyi, A. H., \& Hamden, E. T. 2008, \apj, 676, 1109
\bibitem[Garc\'{\i}a-D\'{\i}az \& Henney(2007)]{gar07} Garc\'{\i}a-D\'{\i}az, Ma.-T., \& Henney, W. J. 2007, \aj, 133, 952 (GD07)
\bibitem[Garc\'{\i}a-D\'{\i}az et al.(2008)]{gar08} Garc\'{\i}a-D\'{\i}az , Ma.-T., Henney, W. J., L\'opez, J. A., \& Doi, T. 2008, \rmxaa, 44, 181
\bibitem[Gaume et al.(1998)]{gau98} Gaume, R. A., Wilson, T. L., Vrba, F. J., Johnston, K. J., \& Schmid-Burgk, J. 1998, \apj, 493, 940
\bibitem[Genzel et al.(1981)]{gen81} Genzel, R., Reid, M. J., Moran, J. M.,\& Downes, D. 1981, \apj, 244, 884
\bibitem[G\'omez et al.(2005)]{gom05} G\'omez, L., Rodr\'{\i}guez, L. F., Loinard, L., Lizano, S., Poveda, A., \& Allen, C. 2005, \apj, 635, 1166
\bibitem[Graham et al.(2003)]{gra03} Graham, M. F., Meaburn, J., \& Redman, M. P. 2003,  \mnras, 343, 419
\bibitem[Hartigan et al.(2001)]{har01} Hartigan, P., Morse, J. A., Reipurth, B., Heathcote, S., \& Bally, J. 2001,\apjl, 559, L157
\bibitem[Hashimoto et al.(2007)]{has07} Hashimoto, J., Tamura, M., Kandori, R., Kusakabe, N., Nakajima, Y., Sato, S., Nagashima, C., Kurita, M., et al. 2007, PASJ, 59, 489
\bibitem[Henney et al.(2007)]{hen07} Henney, W. J., O'Dell, C. R., Zapata, L. A., Garc\'{\i}a-D\'{\i}az, Ma. T., Rodr\'{\i}guez, L. F., \& Robberto, M. 2007, \aj, 133, 2343 (HO07)
\bibitem[Hillenbrand \& Carpenter(2000)]{hil00} Hillenbrand, L. A. , \& Carpenter, J. M. 2000, \apj, 540, 236
\bibitem[Hirota et al.(2007)]{hir07} Hirota, T., Bushimata, T., Choi, Y. K., Honma, M., Imai, H., Iwadate, K., Jike, T. Kameno, S., et al. 2007, PASJ, 59, 897
\bibitem[Hu(1996)]{hu96} Hu, X. 1996, \aj, 112, 2712
\bibitem[Jefferies(2007)]{jef07} Jefferies, R. D. 2007, MNRAS, 376, 1109
\bibitem[Jones \& Walker(1985)]{jw85} Jones, B. F., \& Walker, M. F. 1985, \aj, 90, 1320
\bibitem[Jones \& Walker(1988)]{jw88} Jones, B. F., \& Walker, M. F. 1988, \aj, 95, 1755
\bibitem[Kaifu et al.(2000)]{kai00} Kaify, N., Usuda, T., Hayashi, S. S., Itoh, Y., Akiyama, M., Yamashita, T., Nakajima, Y., Tamura, M., et al. 2000, PASJ, 52, 1
\bibitem[Kraus et al.(2007)]{kra07} Kraus, S., Balega, Y. Y., Berger, J.-P., Hofmann, K.-H., Millan-Gaget, R., Monnier, J. D., Ohnaka, K., Pedretti, E., et al. 2007, \aap, 466, 649
\bibitem[Lee \& Burton(2000)]{lee00} Lee, J.-K., \& Burton, M. G. 2000, \mnras, 315, 11
\bibitem[Meaburn et al.(2003)]{mea03} Meaburn, J., L\'opez, J. A., Guti\'errez, L., Quir\'oz, F., Murillo, J. M., Vald\'ez, J., \& Padrayez, M. 2003, \rmxaa, 39, 185
\bibitem[Menten et al.(2007)]{men07} Menten, K. M., Reid, M. J., Forbrich, J., \& Brunthaler, A. 2007, \aap, 474, 515
\bibitem[Muench et al.(2008)]{mue08} Muench, A., German, K., Hillenbrand, L., \& Preibisch, T. 2008, in Handbook of Star Forming Regions, ed. B. Reipurth, Astron. Soc. Pacific, in press
\bibitem[O'Dell(1998)]{ode98} O'Dell, C. R. 1998, \aj, 115, 263
\bibitem[O'Dell(2001)]{ode01} O'Dell, C. R. 2001, ARAA, 39, 99
\bibitem[O'Dell \& Doi(1999)]{ode99} O'Dell, C. R. \& Doi, T. 1999, \pasp, 111, 1316
\bibitem[O'Dell \& Doi(2003)]{od03} O'Dell, C. R., \& Doi, T. 2003, \aj, 125, 277 (OD03)
\bibitem[O'Dell et al.(1997)]{ode97} O'Dell, C. R., Hartigan, P., Lane, W. M., Wong, S.-K., Burton, M. G., Raymond, J., \& Axon, D. J. 1997, \aj,114, 730
\bibitem[O'Dell et al.(2008)]{ode08} O'Dell, C. R., Muench, A., Smith, N., \& Zapata, L. 2008, in Handbook of Star Forming Regions, ed. B. Reipurth, Astron. Soc. Pacific, in press
\bibitem[O'Dell et al.(2007)]{ode07} O'Dell, C. R., Sabbadin, F., \& Henney, W. J. 2007, \aj, 134, 1679
\bibitem[O'Dell \& Wen(1994)]{ode94} O'Dell, C. R., \& Wen, Z. 1994, \apj, 436, 194 
\bibitem[O'Dell \& Wong(1996)]{ode96} O'Dell, C. R., \& Wong, S.-K. 1996, \aj, 111, 846
\bibitem[O'Dell \& Yusef-Zadeh(2000)]{ode00} O'Dell, C. R., \& Yusef-Zadeh, F.  2000, \aj, 120, 382 
\bibitem[Reipurth(1989)]{rei89} Reipurth, B. 1989, Nature, 340, 42
\bibitem[Reipurth et al.(1997)]{rei97} Reipurth, B., Bally, J., \& Devine, D. 1997, \aj, 114, 2708
\bibitem[Robberto et al.(2005)]{rob05} Robberto, M., Beckwith, S. V. W., Panagia, N., Patel, S. G., Herbst, T. M., Ligori, S., Custo, A., Boccacci, P. \& Bertero, M. 2005, \aj, 129, 1534
\bibitem[Rodr\'{\i}guez et al.(2005)]{rod05} Rodr\'{\i}guez, L. F., Poveda,  A., Lizano, S. \& Allen, C. 2005, \apj, 627, L65
\bibitem[Sandstrom et al.(2007)]{san07} Sandstrom, K. M., Peek, J. E. G., Bolatto, A. D., \& Plambeck, R. L. 2007, \apj, 667, 1161
\bibitem[Schultz et al.(1999)]{schu99} Schultz, A. S. B., Colgan, S. W. J., Erickson, E. F., Kaufman, M. J., \& Hollenbach, D. J. 1999, \apj, 511, 282
\bibitem[Smith et al.(2004)]{smi04} Smith, N., Bally, J., Shuping, R. Y., Morris, M., \& Hayward, T. L. 2004, \apjl, 610, L117
\bibitem[Stanke et al.(2002)]{sta02} Stanke, T., McCaughrean, M. J., \& Zinnecker, H. 2002, \aap\ 392, 239
\bibitem[Takami et al.(2002)]{tak02} Takami, M., Usuda, T., Sugai, J., Suto, H., Pyo, T.-S., Takeyama, N., Aoki, T., Mizutani, K., \& Tanaka, M. 2002, \apj, 566, 910
\bibitem[Walter et al.(1995)]{wal95} Walter, D. K., O'Dell, C. R., Hu, X., \& Dufour, R. J. 1995, \pasp, 107, 686
\bibitem[Wilson et al.(2005)]{wil05} Wilson, B. A., Dame, T. M., Masheder, M. R. W., \& Thaddeus, P. 2005, \aap, 430, 523
\bibitem[Zapata et al.(2006)]{zap06} Zapata, L. A., Ho, P. T. P., Rodr\'{\i}guez, L. F., O'Dell, C. R., Zhang, Q., \& Muench, A. 2006, \apj, 653, 398
\bibitem[Zapata et al.(2004a)]{zap04a} Zapata, L. A., Rodr\'{\i}guez, L. F., Kurtz, S. E., \& O'Dell, C. R., \aj, 127, 2252
\bibitem[Zapata et al.(2004b)]{zap04b} Zapata, L. A., Rodr\'{\i}guez, L. F., Kurtz, S. E., O'Dell, C. R., \& Ho, P. T. P. 2004, \apjl, 610, L121
\bibitem[Zapata et al.(2005)]{zap05} Zapata, L. A., Rodr\'{\i}guez, L. F., Ho. P. T. P., Zhang, Q., Chunhua, Q., \& Kurtz, S. E. 2005, \apjl, 630, L85
\end{thebibliography}
\end{document}